\documentclass[aps,pra,floatfix,footinbib,preprintnumbers,notitlepage,reprint]{revtex4-2}

\usepackage{amsmath,amssymb,mathtools}
\usepackage{bbold}
\usepackage{graphicx}
\usepackage{bm,color}
\usepackage{natbib,hyperref}
\usepackage[capitalise]{cleveref}
\usepackage{enumitem}
\usepackage{adjustbox}
\usepackage{mathrsfs}

\usepackage[normalem]{ulem}
\usepackage{xspace}

\hypersetup{breaklinks=true,colorlinks=true,linkcolor=blue,urlcolor=blue,citecolor=blue}

\newcommand{\be}{\begin{equation}}
\newcommand{\ee}{\end{equation}}
\newcommand{\nn}{\nonumber}
\newcommand{\lf}{\left}
\newcommand{\rt}{\right}

\newcommand{\vphi}{\varphi}

\newcommand{\overbar}[1]{\mkern 1.5mu\overline{\mkern-1.5mu#1\mkern-1.5mu}\mkern 1.5mu}
\newcommand{\latPi}{\overbar{\Pi}}
\newcommand{\latPhi}{\overbar{\Phi}}
\newcommand{\latH}{\overbar{H}}
\newcommand{\latm}{m}
\newcommand{\latlam}{\lambda}
\newcommand{\latf}{f}

\newcommand{\minit}{m_{I}}
\newcommand{\ginit}{\lambda_{I}}
\newcommand{\finit}{f_{I}}

\newcommand{\ket}[1]{\left| #1 \right\rangle}
\newcommand{\bra}[1]{\left\langle #1 \right|}
\DeclarePairedDelimiterX\braket[2]{\langle}{\rangle}{#1 \delimsize\vert #2}
\newcommand{\abs}[1]{\left| #1 \right|}
\newcommand{\sabs}[1]{|#1|}
\newcommand{\sket}[1]{| #1 \rangle}
\newcommand{\ketbra}[2]{\ensuremath{|#1 \vphantom{#2} \rangle \langle #2 \vphantom{#1} | }}
\newcommand{\average}[1]{\ensuremath{\langle#1 \rangle}}
\newcommand{\norm}[1]{\ensuremath{\left| \left|#1 \right| \right|}}
\newcommand{\opmatrix}[3]{\ensuremath{\langle#1 \vphantom{#3} | #2 | #3 \vphantom{#1} \rangle}}

\newcommand{\sandwich}[3]{\lf< #1 \lf| #3 \rt| #2 \rt>}
\newcommand{\ssandwich}[3]{\langle #1 | #3 | #2 \rangle}

\newcommand{\evaE}[2]{E_{#1}(#2)}

\newcommand{\qregsize}{n_{q}}
\newcommand{\ftwo}{\ensuremath{f_{F}}\xspace}
\newcommand{\fzero}{\ensuremath{f_{I}}\xspace}
\newcommand{\fone}{\ensuremath{f_{M}}\xspace}

\newcommand{\oszf}[1]{\sigma^{z}_{#1}}

\def\O{{\cal{O}}}

\def\bt{\bm{\theta}}
\def\tPhi{{\Phi}}
\def\tPi{{\Pi}}
\def\dH{{\cal{H}}}
\def\cH{{\mathscr{H}}}
\def\veps{\varepsilon}
\def\F{{\cal{F}}}

\def\gZZ{~\mathclap{Z}Z}
\def\gZZZZ{~\mathclap{Z}Z\mathclap{Z}Z}

\newcommand{\edit}[1]{#1}

\begin{document}
\preprint{FERMILAB-PUB-22-757-QIS}
\title{Simulating scalar field theories on quantum computers with limited resources}
\author{Andy C.~Y.~Li, Alexandru Macridin, Stephen Mrenna, Panagiotis Spentzouris}
\affiliation{Fermi National Accelerator Laboratory, Batavia, IL 60510}
\date{\today}

\begin{abstract}
We present a quantum algorithm for implementing $\phi^4$ lattice scalar field theory on qubit computers. The field is represented in the discretized field amplitude basis. The number of  qubits and elementary gates required by the implementation  of the evolution operator is proportional to the lattice size. The algorithm allows efficient $\phi^4$ state preparation for a large range of input parameters in both the normal and broken-symmetry phases. The states are prepared using a combination of variational and adiabatic evolution methods. First, the ground state of a local Hamiltonian, which includes the $\phi^4$ self-interaction, is prepared using short variational circuits. Next, this state is evolved by switching on  the coupling between the lattice sites adiabatically. The parameters defining the local Hamiltonian are adjustable and constitute the input of our algorithm. We present a method to optimize these parameters in order to reduce the adiabatic time required for state preparation. For preparing broken-symmetry states, the adiabatic evolution problems caused by crossing the phase transition critical line and by the degeneracy of the broken-symmetry ground state can be addressed using an auxiliary external field which gradually turns off during the adiabatic process. We show that the time dependence of the external field during the adiabatic evolution is important for addressing the broken-symmetry ground state degeneracy. The adiabatic time dependence on the inverse error tolerance can be reduced from quadratic to linear by using a field strength that decreases exponentially in time relative to one that decreases linearly.
\end{abstract}

\maketitle

\section{Introduction}
Simulating highly entangled quantum systems is among the first applications of quantum computers expected to show a practical advantage over classical computers \cite{Preskill2018,Daley2022}. 
The development and application of new quantum processors \cite{Reagor2018,Arute2019,Jurcevic2021} may allow for revolutionary calculations in quantum chemistry \cite{O'Malley2016,Kandala2017,Hempel2018,Colless2018,Kandala2019,Google2020}, condensed-matter physics \cite{macridin_prl_2018,macridin_pra_2018,Gluza2018,Ma2020,Uvarov2020,Ji2020}, nuclear physics \cite{Roggero2020,Roggero2020b} and high-energy physics \cite{Jordan2012,Klco2018,Pedernales2018,Hu2019,Lamm2019}. 
The simulation of relativistic quantum field theory on quantum hardware \cite{Jordan2012,Klco2018,Lamm2019,farrelly_2020,Klco2020,Barata2021} has been an active research topic in recent years. In practice, the mapping and preparation of continuous fields on  near-future quantum hardware of limited size and with limited control fidelity provides a number of  challenges. In this paper, we address the simulation of the $\phi^4$  scalar field on digital quantum computers.

The $\phi^4$  scalar field model~\cite{Aizenman1981,Frohlich1982} is  a simplified model of the Higgs field of the standard model of particle physics and has been intensively studied over the years. 
Despite its apparent simplicity, it has rich physics. For example, in $(1+1)$ and $(2+1)$ space-time dimensions, it exhibits a phase transition to a broken-symmetry phase characterized by a finite vacuum expectation value $\average{\phi}$ \cite{Chang1976,Magruder1976}.
Perturbative methods based on a diagrammatic expansion are only valid in the weak interaction regime.  As a result, the strong interacting regime has been studied numerically.
Calculations of the critical coupling and exponent for $(1+1)$ dimensions have been performed using DMRG~\cite{Sugihara_2004}, tensor network methods~\cite{milsted_prd_2013,Banuls2020}, Monte Carlo methods~\cite{Schaich2009,Bosetti2015} and diagonaliztion methods~\cite{LEE2001223,Rychkov_prd_2015}.
However, since the Hilbert space of the $\phi^4$ model is exponentially large, the  field degrees of freedom 
must be truncated,  making the extrapolation of the  numerical results to the continuous limit challenging and not always reliable.

Quantum simulations can overcome the size problems related to the Hilbert space and, unlike most classical Monte Carlo methods, can calculate the real-time correlations and  nonequilibrium dynamics of the system. The bosonic fields can be represented efficiently on qubits in a discretized field amplitude basis \cite{macridin_pra_2018,macridin_prl_2018,Macridin2021}. However the preparation of field eigenstates on qubits is not straightforward.  For example, the method proposed in Refs~\cite{Jordan2012,Klco2020,Barata2021} prepares an initial noninteracting multivariate Gaussian state and uses adiabatic continuation to reach the desired interacting state.
However, the construction of a multivariate Gaussian wave function using the Kitaev-Webb method~\cite{Kitaev2008} requires a very large number of qubits   and is not feasible on near-term quantum hardware.
Moreover, the  preparation of broken-symmetry states using adiabatic continuation of noninteracting states is challenging since the adiabatic path has to cross a critical region with a vanishing excitation gap.   Furthermore, the ground state of broken-symmetry states is degenerate, causing further complications.

Here, we present a quantum algorithm for lattice $\phi^4$ field evolution on qubits and a method for initial state preparation suitable for near-term quantum computers. A relatively small number of qubits per lattice site, $\qregsize \approx 6 \sim 8$,  is sufficient to address even strong-coupling regimes.
The number of qubits and the number of gates scale proportionally to the system size $N$.
The most expensive part of the algorithm arises from the implementation of the $\phi^4$ interaction, which requires $\mathcal{O}(N \, \qregsize^4)$ two-qubit gates, while 
the implementation of the other terms in the Hamiltonian requires  $\mathcal{O}(N \, \qregsize^2)$ two-qubit gates. The field state preparation combines variational and adiabatic evolution approaches. 
The Hamiltonian is split into two parts, a local one that sums contributions from each individual site and a nonlocal one containing coupling between sites.  The adiabatic process starts from the
ground state of the local Hamiltonian. Then the coupling between sites is \edit{turned} on adiabatically. 
Unlike previous approaches in the literature \cite{Jordan2012}, our method introduces self-interactions from the start. The ground state of the local Hamiltonian is prepared accurately using short variational circuits.
Instead of preparing the full lattice states using variational ansatzes \cite{Liu2021}, which would be difficult to scale up due to Barren plateaus \cite{McClean2018}, our variational circuits prepare local states. The calculation of these circuits' parameters is independent of the system size and can be done easily on classical computers using various optimization methods.
The input parameters of the local Hamiltonian can be adjusted to minimize 
the time of the adiabatic process. We find a direct correlation between the adiabatic time and the 
local overlap of the initial wave function and the target wave function. 
We propose a strategy to determine the optimal parameters of the local Hamiltonian
by maximizing this local overlap.

We also address the problems associated with the preparation of the broken-symmetry states, namely
the crossing of the critical phase transition region characterized by a vanishing excitation gap and
the double degeneracy of the broken-symmetry state. Both of these problems can be mitigated by
coupling the scalar field to an external field. We propose a two step adiabatic process for preparing
broken-symmetry states. The first adiabatic process starts from a local state in the presence of a significant external field. Then, adiabatically,  the inter-site coupling term is turned on and the external field is decreased.  Due to the presence of the external field, the excitation gap is robust during this process. The second adiabatic process starts from the terminus of the first  one. During this step the external field is decreased to vanishing values. 
We find an reduction of the required adiabatic time from $\mathcal{O} (\veps^{-2})$ to $\mathcal{O} (\veps^{-1}\ln(\veps^{-1}))$ with $\veps$ being the error bound when the external field  decreases exponentially in time compared to the case of linear decrease in time.

This paper is organized as follows. We review the $\phi^4$ model and its lattice discretization in \cref{sec:model}. We then discuss the qubit encoding and circuits to simulate the scalar field evolution on quantum computers in \cref{sec:field}. 
In \cref{sec:state_preparation}, we introduce our state preparation protocol consisting of the variational local-state preparation (\cref{subsec:var_prep}), and adiabatic evolution for the normal phase (\cref{sssec:ad_normalphase}) and for the broken-symmetry phase (\cref{ssec:adiabatic_prep_bs}) supported by numerical simulation of the lattice $\phi^4$ model with up to four sites.  Our summary and conclusions are provided in \cref{sec:conclusion}.

\section{The \texorpdfstring{$\phi^4$}{phi4} model}
\label{sec:model}
\begin{figure}[tb]
	\begin{center}
		\includegraphics*[width=1.0\linewidth]{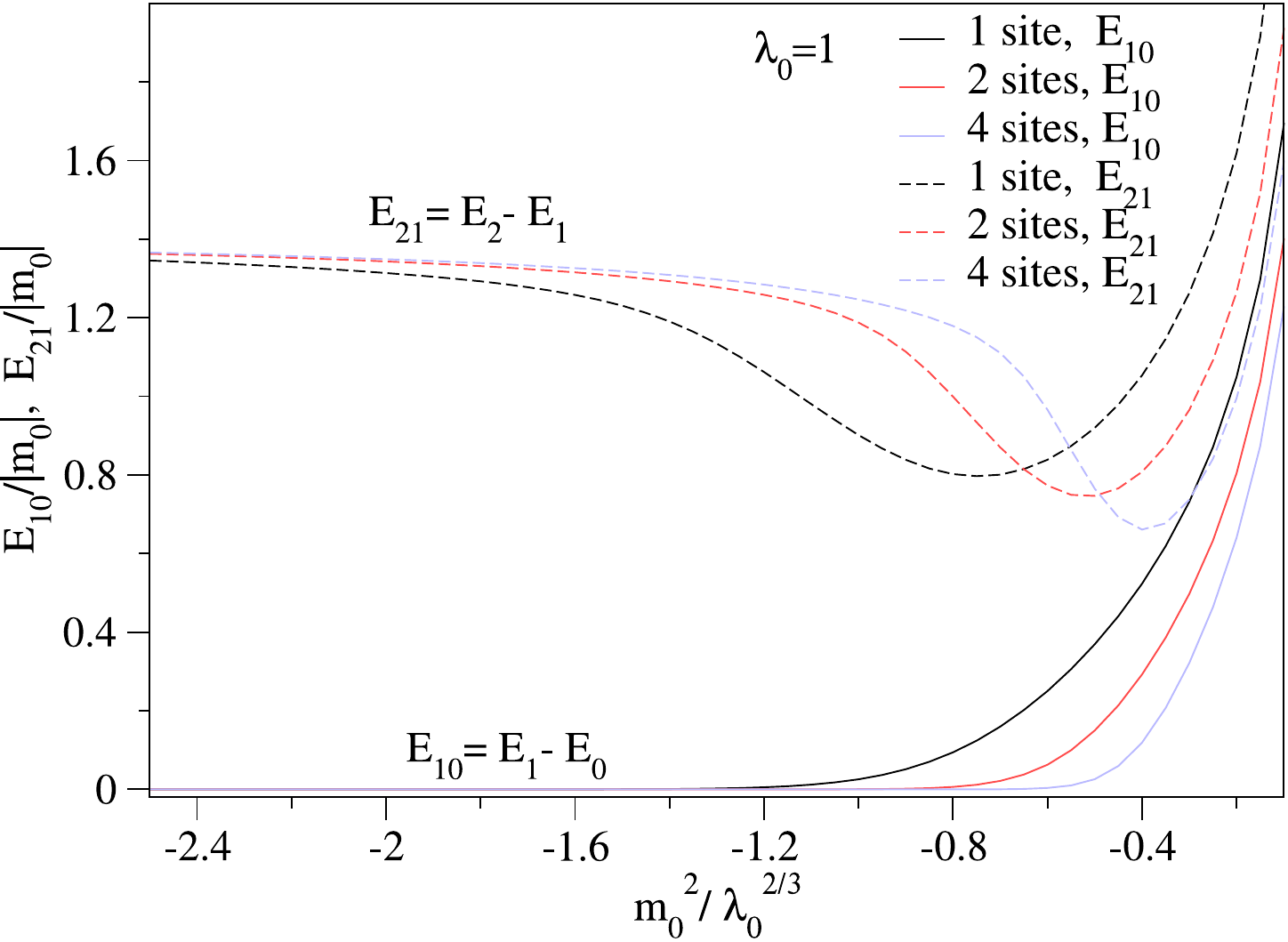}
	\end{center}
	\caption{
		The gap, $E_{10}=E_1-E_0$, and the energy difference between the second and the first excited states, $E_{21}=E_2-E_1$, for one-site, two-site and four-site $\phi^4$ lattices for negative values of $\latm_0^2$.
		The system is nearly double degenerate when $\latm_0^2/\latlam_0^{2/3} \ll -1$. The gap decreases exponentially with increasing the magnitude of $|\latm_0^2|$. On the other hand, $E_{21}$ increases slightly with increasing the magnitude of $|\latm_0^2|$. With increasing the number of sites, the gap $E_{10}$ decreases while $E_{21}$ increases.}
	\label{fig:gaps14}
\end{figure}

The Hamiltonian density of the $\phi^4$ model can be written as ($\hbar = c = 1$)
\begin{align}
	\label{eq:Hcont_def}
	\mathbb{H} &=
	\frac{1}{2} \pi^2
	+\frac{1}{2} m_{b}^2 \phi^2
	+\frac{1}{2} \lf(\nabla \phi \rt)^2
	+\frac{\lambda_b}{4!}\phi^4 + f_b \phi
	,
\end{align}
where $m_b$ and $\lambda_b$ are the unrenormalized (bare) mass and interaction strength, respectively. In order to investigate the broken-symmetry phase, it is convenient to consider a coupling term between the scalar field and a external field $f_b$. The field operator $\phi$ and the conjugate-field operator $\pi$ obey the commutation relation,
\begin{align}
	\label{eq:com_cont}
	\lf[ \pi(x), \phi(y)\rt]=i\delta(x-y).
\end{align}

For quantum simulation, we consider the lattice version of the $\phi^4$ model in $d+1$ spacetime dimensions given by
\begin{align}
	\label{eq:Hdis}
	H_{\mathrm{lat}}  =  a^d \sum_{j} \Bigg[ &
	\frac{1}{2} \pi_{j}^2
	+ \frac{1}{2} m_b^2 \phi_{j}^2
	+ \frac{1}{2a^2} \sum_{e=1}^d\lf( \phi_{j+e}-\phi_j \rt)^2
	\nn\\
	&
	+ \frac{\lambda_b}{4!} \phi_{j}^4 + f_b \phi_j
	\Bigg]
\end{align}
where $a$ is the lattice constant and $j$ labels lattice sites. The label  $j+e$ represents the next-nearest neighbor site of the site $j$ in the direction $e$.
Note that \edit{in ~\cref{eq:Hdis}} only the space dimension is discretized and on a lattice. This differs from most \edit{models employed} in Monte Carlo-based simulations, where both space and imaginary time dimensions are discretized \edit{on a lattice}.
\edit{Nonetheless, the implementation of our algorithm on quantum computers also requires time discretization, consequence of the Trotter-Suzuki expansion~\cite{trotter_1959,suzuki_1976,Lloyd1996}
of the time evolution operator.}
The lattice field operators in \cref{eq:Hdis} obey the commutation relations
\be
\lf[\phi_{i}, \pi_{j}\rt] =  i a^{-d} \delta_{i, j}
\text{\ \ and \ \ }
\lf[\phi_{i}, \phi_{j}\rt] =  \lf[\pi_{i}, \pi_{j}\rt] = 0.
\ee
The continuous limit is achieved by taking $a \rightarrow 0$.
It is convenient to introduce dimensionless field variables
\begin{align}
	\label{eq:nf}
	\latPhi_{j} & = a^{\frac{d -1}{2}} \phi_{j}
	\text{\ \ and \ \ }
	\latPi_{j} =  a^{\frac{d +1}{2}} \pi_{j}
\end{align}
which obey the canonical commutation relations
\begin{align}
	\label{eq:fp_comm}
	\lf[\latPhi_{i}, \latPi_{j}\rt] & =  i \delta_{i, j}
	\text{\ \ and \ \ }
	\lf[\latPhi_{i}, \latPhi_{j}\rt]  =  \lf[\latPi_i, \latPi_j\rt] =  0.
\end{align}
Using these dimensionless operators, the Hamiltonian is given by
\begin{align}
	\label{eq:Hdis1}
	\latH  =
	\sum_{j} \Bigg[&
	\frac{1}{2} \latPi_{j}^2 + \frac{1}{2}  \latm_0^2  \latPhi_{j}^2
	+\frac{1}{2}\sum_{e=1}^{d} \lf( \latPhi_{j+e } - \latPhi_{j}\rt)^2
	\nn\\ 
	&  + \frac{\latlam_0}{ 4!} \latPhi_{j}^4 + \latf_0 \latPhi_{j} 
	\Bigg].
\end{align}
where $\latH \equiv H_{\mathrm{lat}}a$, $\latm_0^2 \equiv m_b^2 a^2$, $\latlam_0 \equiv \lambda_b a^{3-d}$ and $\latf_0 \equiv f_b a^{\left(3+d\right)/2}$ are dimensionless.
This Hamiltonian (with $\latf_0=0$) was previously used in numerical simulations of the scalar field model \cite{milsted_prd_2013,farrelly_2020}. It represents a set of coupled harmonic oscillators with an anharmonic interaction.

The correlation length is a measurable parameter 
which determines how the correlation between the field values at two separate points decays with the distance between these points. To extrapolate the lattice results to a meaningful continuous limit with a finite correlation length, it is necessary to simulate large lattices for Hamiltonian parameters corresponding to large lattice correlation lengths (measured in units of $a$). The physics of the $\phi^4$ theory is extracted from the lattice results by taking  $a/\xi \rightarrow 0$ (continuous limit) and $L/\xi \rightarrow \infty$ (macroscopic limit), where $L$ is the lattice size and $\xi$ is the lattice correlation length. Equivalently, this implies simulations with $m_{p} a \rightarrow 0$ and  $m_{p} a \gg 1/N$ where $N$ is the number of lattice sites per dimension and $m_p \propto 1/\xi$ is the physical mass.

While the lattice physical parameters are needed for the extrapolation from the lattice to the continuous theory, the bare parameters define the input of the simulations. To be useful, a quantum algorithm should be able to prepare efficiently quantum states for a large range of bare parameters, including both negative and positive input parameter $m_0^2$.
In principle, the relation between the physical and the bare parameters can be established from simulations, since the lattice physical parameters can be extracted from the correlation functions. 
For the extrapolation to the continuous limit the bare lattice parameters need to be chosen dependent on the lattice constant $a$. This dependence is significant since the renormalization theory shows that, in order to extrapolate to a continuous theory with finite physical observables, the bare parameters diverge with $a\rightarrow 0$ in many cases. For example, in $(1+1)$ and $(2+1)$ dimensions, the bare squared mass $m_0^2$ becomes negative and proportional to $\ln(m_p a)$ and $-1/\left(m_p a\right)$, respectively \cite{jordan_qic_2014}, for small $a$.

The $\phi^4$ model (with $f_0=0$) has a discrete $Z_2$ symmetry from the transformation $\phi \rightarrow -\phi$. In $(3+1)$ dimensions, the theory is "believed" to be trivial (\textit{i.e.} the theory is actually noninteracting in the continuum limit), although no rigorous proof exists~\cite{luscher_NPB_1987,FREEDMAN1982481,FOX1985148,lang1985block,DRUMMOND198725}. For higher dimensions, the triviality can be rigorously proven~\cite{Aizenman1981}.
In $(1+1)$ and $(2+1)$ dimensions, the model exhibits a phase transition from a symmetric state with $\average{\phi}=0$ to a broken-symmetry phase with finite $\average{\phi}$ \cite{Chang1976,Magruder1976}.
However, in finite size systems, like the ones used for simulations, the ground state cannot have a broken symmetry and there is no phase transition. Nonetheless, the broken-symmetry phase can be investigated numerically by considering the coupling term  $\latf_0 \latPhi$ in \cref{eq:Hdis1} and extrapolating the results to the large lattice size ($L \rightarrow \infty$) and the zero external field ($\latf_0 \rightarrow 0$) limits. 

It is interesting that, for negative values of $\latm_0^2$ and small interaction strength ({\emph{i.e.}} when $|\latm_0|^3/\latlam_0 \gg 1$), the ground state is nearly twofold degenerate and exhibits properties characteristic of the broken-symmetry phase even for small lattices.  A single site system reduces to a double-well potential Hamiltonian for negative $\latm_0^2$.
The field distribution in the ground state is symmetric and double-peaked around zero, with the two maxima located at the points $\pm \latPhi_{\text{m}}$ which minimize the potential. The gap is small, decreasing exponentially fast with increasing magnitude of $|\latm_0|^3/\latlam_0$.
A small external field $\latf_0$ (of the order of the gap) coupled to the scalar field amplitude produces a ground state with  finite $\average{\latPhi} \approx \latPhi_{\text{m}}$ (or $\average{\latPhi} \approx -\latPhi_{\text{m}}$, depending on $\latf_0$ sign).  Numerical calculations of small size systems show that the system remains nearly twofold degenerate when the number of sites is increased. The gap decreases with an increasing number of sites, while the energy difference between the second and the first excited states does not decrease, as can be seen in \cref{fig:gaps14}.
This is a consequence of the kinetic term in the Hamiltonian (the third term in \cref{eq:Hdis1}) which favors similar field configurations at neighboring sites. These properties of small size systems allows us to investigate quantum state preparation methods for broken-symmetry phase by using classical simulations of small lattices, as discussed in \cref{ssec:adiabatic_prep_bs}.

\section{\texorpdfstring{$\Phi^4$}~    field on qubits}
\label{sec:field}
This section describes the qubit encoding of the bosonic states and the implementation of the  evolution operator corresponding to the lattice Hamiltonian $H_{\mathrm{lat}}$ in \cref{eq:Hdis1}.
Discretizing continuous groups is important for simulating quantum field theories and has been studied for several models \cite{Ciavarella2021,Alam2022,Gustafson2022}.
The representation of bosonic fields on qubits was discussed in detail in \cite{Macridin2021}. We will review briefly the general methodology in \cref{ssec:repr}, present the qubit encoding in \cref{ssec:qubitencoding}, and address the evolution operator implementation in \cref{ssec:trotter}.

\subsection{Finite representation of bosonic fields}
\label{ssec:repr}

The lattice Hilbert space is a direct product of local Hilbert spaces (one at each lattice site) such that $\cH=\prod_{j=1}^N \otimes \cH_j$, where $j$ labels the lattice site and $N$ is the number of lattice sites. A local Hilbert space $\cH_j$ is infinite dimensional. The number of bosons contributing to the wave function is, in principle, unbounded.
However, since we are interested in the low-energy physics of the system, we postulate that, at every lattice site, the number of bosons can be truncated with negligible error by a cutoff number $N_b$. 

The eigenvectors $\{\ket{\vphi}_j\}$ of the field operator,
\be
\label{eq:phieigen}
\latPhi_j\ket{\vphi}_j=\vphi\ket{\vphi}_j,
\ee
form a convenient basis choice for representing the evolution operator since the Hamiltonian interaction terms are diagonal in this basis.
However, the eigenvalues $\vphi \in \mathbb{R}$ are continuous and unbounded. Therefore, discretization procedures are necessary to represent the truncated Hilbert space in the field amplitude basis. We introduce the discretization procedure below.

The low-energy subspace of the local Hilbert space $\cH_j$ is spanned by the states with a number of bosons below the cutoff $N_b$ and can be represented with good accuracy by a finite Hilbert space $\dH_j$ of dimension $N_\vphi$, with  $N_\vphi > N_b$, as described below. 
Let  $\{\ket{\vphi_\alpha}_j\}$  be a set of orthonormal vectors belonging to $\dH_j$ with \edit{ $\alpha \in \{0,1,\cdots,N_{\vphi}-1\}$ }.
We define the \emph{discrete field operators} $\tPhi_j$ and  $\tPi_j$ acting on $\dH_j$  as
\begin{align}
	\label{eq:phiopdef}
	\tPhi_j\ket{\vphi_\alpha}_j&=\vphi_\alpha\ket{\vphi_\alpha}_j, \\
	\label{eq:piopdef}
	\tPi_j& = \mu \F_j \tPhi_j \F_j^{-1}.
\end{align}
where $\vphi_\alpha$ is the discrete eigenvalue,
\begin{align}
	\label{eq:vphialpha}
	\vphi_\alpha &= \Delta_\vphi \left(\alpha-\frac{N_\vphi-1}{2}\right),~\alpha \in \{0,1,\cdots,N_{\vphi}-1\}
	\\
	\label{eq:deltadef}
	\Delta_\vphi&=\sqrt{\frac{2 \pi }{N_\vphi \mu}}.
\end{align}
and $\F_j$ is the finite Fourier transform,
\begin{align}
	\label{eq:fft}
	\F_j =\frac{1}{\sqrt{N_\vphi}}\sum_{\alpha,\beta=0}^{N_{\vphi}-1} e^{i \frac{2 \pi}{N_\vphi} \left(\alpha-\frac{N_{\vphi}-1}{2}\right) \left(\beta-\frac{N_{\vphi}-1}{2}\right)}\ket{\vphi_\alpha}_{j}\bra{\vphi_\beta}_j.
\end{align}
In \cref{eq:piopdef}, $\mu>0$ is the boson mass which is the parameter entering in the definition of the lattice boson creation and annihilation operators,
\begin{align}
	a_j^{\dagger}=\sqrt{\frac{\mu}{2}}\latPhi_j-i\sqrt{\frac{1}{2\mu}}\latPi_j, ~~ a_j=\sqrt{\frac{\mu}{2}}\latPhi_j+i \sqrt{\frac{1}{2\mu}}\latPi_j.
\end{align}

The definition of  $\tPhi_j$ given by \cref{eq:phiopdef,eq:vphialpha,eq:deltadef} represents the discretized and truncated version of \cref{eq:phieigen}.
The set of states \edit{ $\{\ket{\kappa_\beta}_j\}_{\beta \in \{0,1,\cdots,N_\vphi-1\}}$ } obtained by applying the Fourier transform to the set $\{\ket{\vphi_\alpha}_j\}$ w,
\begin{align}
	\label{eq:kappasatate}
	\ket{\kappa_\beta}_j &\equiv \F_j \ket{\vphi_\beta}_j \\ \nonumber &=\frac{1}{\sqrt{N_\vphi}}\sum_{\alpha=0}^{N_{\vphi}-1} e^{i \frac{2 \pi}{N_\vphi} \left(\alpha-\frac{N_{\vphi}-1}{2}\right) \left(\beta-\frac{N_{\vphi}-1}{2}\right)}\ket{\vphi_\alpha}_j
\end{align}
are the eigenvectors of the discrete conjugate-field  operator $\tPi_j$ defined by \cref{eq:piopdef} such that
\begin{align}
	\label{eq:phopeigen}
	\tPi_j\ket{\kappa_\beta}_j=\kappa_\beta\ket{\kappa_\beta}_j
\end{align}   
where
\begin{align}
	\label{eq:deltak1}
	\kappa_\beta&=\Delta_\kappa \left(\beta-\frac{N_\vphi-1}{2}\right),~\beta \in \{0,1,\cdots,N_{\vphi}-1\}
	\\
	\label{eq:deltak2}
	\Delta_\kappa&=\sqrt{\frac{2 \pi \mu}{N_\vphi}}.
\end{align}
\Cref{eq:phopeigen} is the discretized version of the conjugate-field operator eigenvalue equation, $\latPi_j\ket{\kappa}_j=\kappa\ket{\kappa}_j$, with continuous and unbounded $\kappa \in \mathbb{R}$.

Different representations corresponding to different values of $\mu$ can be chosen to construct the finite representation.
For a given problem and desired accuracy, the cutoff $N_b$ depends on the boson mass $\mu$. In principle $\mu$ should be optimized for the lowest possible  cutoff $N_b$ to reduce the computing resources. 
Moreover, as can be seen from \cref{eq:deltadef,eq:deltak2}, the discretization interval $\Delta_\vphi$ of the field amplitude variable and the discretization interval $\Delta_\kappa$ of the conjugate-field variable are also dependent on the boson mass parameter $\mu$.  The parameter $\mu$ can be tuned to adjust the accuracy of the discretization. Increasing $\mu$ decreases the field variable discretization interval and increases the conjugate-field discretization interval. The discretized field and conjugate-field variables are related by a finite Fourier transform, thus $\Delta_\vphi \Delta_\kappa =2 \pi /N_\vphi$. To decrease both discretization intervals, $\Delta_\vphi$ and  $\Delta_\kappa$, the number of discretization points $N_\vphi$ should be increased. For quantum simulations, tuning $\mu$ to increase the accuracy of the wave function's discretization is much easier than the process of optimizing $\mu$ to decrease the boson number cutoff $N_b$, as discussed in \cite{Macridin2021}.

On the subspace of $\dH_j$ spanned by the first $N_b$ eigenstates of the harmonic oscillator Hamiltonian ($H_{hj}=\frac{1}{2}\tPi_j^2+\frac{1}{2}\mu^2\tPhi_j^2$), the  discrete field and conjugate-field operators obey, with $\O(\epsilon)$ accuracy, the canonical
commutation relation, 
\be
I_{Nb}\left[\tPhi_j,\tPi_j\right]I_{Nb}=iI_{Nb}+\O(\epsilon).
\ee
Here, $I_{Nb}$ is the projector on the $N_b$ size low-energy subspace of the harmonic oscillator. This is a consequence of the Nyquist-Shannon sampling theorem applied to the fast decaying boson number wave functions, as discussed in \cite{Macridin2021}.
For a problem of interest, as long as $N_b$ is taken large enough such that the contribution of states with more than $N_b$ bosons can be neglected, the infinite Hilbert space $\cH_j$ can be replaced by the finite $N_\vphi$-size Hilbert space $\dH_j$, and the lattice field operators $\latPhi_j$ and $\latPi_j$ [\cref{eq:nf}] can be replaced by the discrete operators $\tPhi_j$ and $\tPi_j$ [\cref{eq:phiopdef,eq:piopdef}] with $\O(\epsilon)$ accuracy. For a fixed $N_b$, the error $\O(\epsilon)$ decreases exponentially by increasing $N_\vphi$. For practical purpose, we find numerically that a number of discretization points $N_\vphi=2N_b$ yields an accuracy of order $10^{-4}$.

The finite lattice representation is given by the finite Hilbert space $\dH=\prod_{j=1}^N \otimes \dH_j$ of dimension $N_\vphi^N$ and the set of local field and conjugate field-operators \edit{$\{\tPhi_j\}_{j \in \{1, 2, \cdots,N\}}$ and $\{\tPi_j\}_{j \in \{1,2, \cdots, N\}}$} defined by \cref{eq:phiopdef} and \cref{eq:piopdef}, respectively.
The discretized field amplitude basis vectors are
\begin{align}
	\label{eq:latt_state}
	\ket{\vphi_\alpha} \equiv \ket{\vphi_{\alpha 1}}_1\ket{\vphi_{\alpha 2}}_2...\ket{\vphi_{\alpha N}}_N
\end{align}
where
\begin{align}
	\label{eq:alpha}
	\alpha = \{ \alpha_1,\alpha_2,...,\alpha_N \} ~~\text{with}~~\alpha_j \in \{0,1,...,N_\vphi-1\}.
\end{align}

\subsection{Qubit encoding of the finite representation}
\label{ssec:qubitencoding}

The discretized field amplitude basis $\{\ket{\vphi_\alpha}\}$ [\cref{eq:latt_state}] can be encoded on qubits using the binary representation of the label $\alpha$ [\cref{eq:alpha}]. For each site, a register of $n_q=\log_2(N_\vphi)$ qubits is assigned. A local field amplitude state $\ket{\vphi_{\alpha_j}}_j$ at site $j$ is encoded as
\begin{align}
	\label{eq:qbenc1}
	\ket{\vphi_{\alpha_j}}_j \equiv \ket{\alpha_{0j}}_j \ket{\alpha_{1j}}_j...\ket{\alpha_{\left(n_q-1\right)j}}_j
\end{align}
where $\ket{\alpha_{qj}}_j \in \{\ket{0}, \ket{1} \}$ is the $q$-th qubit-state from the register $j$ ({\em i.e.} allocated to represent the field at the site $j$) such that 
\begin{align}
	\label{eq:binary}
	\alpha_j = \sum_{q=0}^{n_q-1} \alpha_{qj} 2^{n_q-1-q}.
\end{align}
Note that the binary variables $\alpha_{qj} \in \{0,1\}$ defined by \cref{eq:binary} yield the binary representation of the integer $ \alpha_j \equiv [\alpha_{0j} \alpha_{1j}...\alpha_{\left(n_q-1\right)j}] $. A lattice state [\cref{eq:latt_state}] is encoded as a direct product of $N$ local states encoded by \cref{eq:qbenc1}. The  lattice states require $N \log_2(N_\vphi)$ qubits for encoding.

The discrete field operator $\tPhi_j$ acting on the $n_q$ qubits assigned to encode the field at site $j$ can be written as
\begin{align}
	\label{eq:phiqub}
	\tPhi_j &= -\Delta_\vphi \sum_{q=0}^{\qregsize - 1} 2^{\qregsize - 1 - q} \frac{\oszf{qj}}{2}
\end{align}
where $\oszf{qj} = \ketbra{0}{0}_{qj} - \ketbra{1}{1}_{qj}$ is the Pauli $Z$ operator and $q$ is the qubit index.
It can be directly checked that  $\tPhi_j$ defined here and the vector encoded as in \cref{eq:qbenc1} satisfy the eigenvalue equation defined by \cref{eq:phiopdef,eq:vphialpha}.

The definition of the conjugate-field operator $\tPi_j$ on the qubit space requires first the qubit implementation of the Fourier transform $\F_j$ [see \cref{eq:piopdef}]. The implementation of Quantum Fourier transform (QFT) on qubits is well known~\cite{nielsen2002quantum}. However, the Fourier transform $\F_j$ defined by \cref{eq:fft} is centered, {\em i.e.} the summation index runs from $-(N_\vphi-1)/2$ to $(N_\vphi-1)/2$, unlike the off-centered QFT where the summation index runs from $0$ to $N_\vphi-1$. As shown in \cref{app:ft_gate}, the Fourier transform is related to the QFT by
\begin{align}
	\label{eq:fftqft}
	\F_j = & e^{-i\frac{ N_\vphi \delta^2 }{2 \pi}} \prod_{q=0}^{n_q-1}R^z_{qj}\left(2^{n_q-1-q} \delta\right) ~\text{QFT}_j
	\nn\\
	& \times \prod_{q=0}^{n_q-1} R^z_{qj}\left(2^{n_q-1-q}\delta \right)
\end{align}
where $\delta=\pi \frac{N_\vphi-1}{N_\vphi}$ and $R^z_{qj}$ is a single-qubit $z$ rotation acting on the qubit $q$ at site $j$ given by
\begin{align}
	\label{eq:rzdef}
	R^z_{qj}(\theta)\equiv e^{-i \theta \frac{\sigma^z_{qj}}{2}}=
	e^{-i\frac{\theta}{2}} \ketbra{0}{0}_{qj} + e^{i \frac{\theta}{2}} \ketbra{1}{1}_{qj}.
\end{align}

According to \cref{eq:piopdef}, the discrete conjugate-field operator is 
\begin{align}
	\label{eq:piqub}
	\tPi_j= \F_j \left(-\Delta_k  \sum_{q=0}^{\qregsize - 1} 2^{q} \frac{ \oszf{qj}}{2}\right)  \F_j^{-1}.
\end{align}
Note that the factor before Pauli $\sigma^z_{qj}$ gate  is $2^q$, unlike the factor in \cref{eq:phiqub} which is $2^{n_q-1-q}$. This is caused by the fact that the qubit order is reversed after a QFT gate (unless additional swap operations are performed to manually reverse the qubit order) \cite{nielsen2002quantum}.

\subsection{Evolution operator}
\label{ssec:trotter}
In order to implement the evolution operator we employ the Trotter-Suzuki expansion \cite{trotter_1959,suzuki_1976,Lloyd1996}. The evolution operator is written as a product of short-time evolution operators corresponding to the different terms in the Hamiltonian, called Trotter steps. Here, we present the qubit implementation of the Trotter steps corresponding to the  different terms present in the $\phi^4$ Hamiltonian. 

We start with the operator $e^{-i \theta \Phi_j}$, where $\theta$ is the time interval of the Trotter step. This Trotter step implements the evolution of the term $\latf_0\Phi_j$ in \cref{eq:Hdis1}. Employing \cref{eq:phiqub} one has
\begin{align}
	\label{eq:trdispl}
	e^{-i \theta \tPhi_j}
	=\prod_{q=0}^{n_q-1} R^z_{qj}\left(- 2^{n_q-1-q} \Delta_\vphi \theta \right).
\end{align}
It reduces to $n_q$ single-qubit $z$ rotations. 

The Trotter step $e^{-i \theta \tPhi_j^2}$ can be written as
\begin{align}
	\label{eq:trphi2}
	e^{-i \theta \tPhi_j^2}=
	e^{-i \theta \Delta_\vphi^2 \frac{N_\vphi^2-1}{12} } \prod_{p=0}^{n_q-1}\prod_{q=0}^{p-1} \gZZ_{pj;qj} \left( \nu_{pq} \right)
\end{align}
where 
\begin{align}
	\gZZ_{pj;qj}(\nu) = & \, e^{-i \nu \sigma^z_{pj} \sigma^z_{qj}}
	\label{eq:nu}
	\\
	\nu_{pq} = & \, 2^{2n_q-3-p-q} \Delta_\vphi^2 \theta.
\end{align}
The two-qubit gate  $\gZZ_{pj;qj}$ acts on the qubit $p$ at site $j$ and on the qubit $q$ at site $j$. Note that one $\gZZ_{pj;qj}$ can be decomposed into two CNOT gates and one $R^z$ gate \cite{Welch2014}. Hence, the Trotter step \cref{eq:trphi2} consists of $n_q(n_q-1)$ CNOT gates.

The implementation of the Trotter step $e^{-i \theta \tPi_j^2}$ is given by
\begin{align}
	\label{eq:trpi2}
	e^{-i \theta \tPi_j^2}&=\F_j e^{-i \mu \theta \tPhi_j^2} \F_j^{-1} \nn\\
	&=
	e^{-i \theta \Delta_\kappa^2 \frac{N_\vphi^2-1}{12} } 
	\F_j \left[\prod_{p=0}^{n_q-1}\prod_{q=0}^{p-1} \gZZ_{pj;qj} \left( \nu'_{pq} \right)\right]\F_j^{-1} 
\end{align}
where $\nu'_{pq}=2^{p+q-1}\theta \Delta_\kappa^2$.
The $\gZZ$ gate's parameter $\nu'_{pq}$ entering in \cref{eq:trpi2} can be obtained from \cref{eq:nu}  by replacing  $n_q-1-p \longrightarrow p$ and $n_q-1-q \longrightarrow q$, (consequence of reverse qubit order after applying QFT) and $\Delta_\vphi \longrightarrow \Delta_\kappa $.
Since  QFT requires $n_q(n_q-1)/2$ CNOT gates, this Trotter step consists of $3n_q(n_q-1)$ CNOT gates.

The Trotter step $e^{-i \theta \tPhi_j \tPhi_l }$ corresponding to the coupling term between the sites $j$ and $l$ is
\begin{align}
	\label{eq:trphijphik}
	e^{-i \theta \tPhi_j \tPhi_l}=\prod_{p=0}^{n_q-1}\prod_{q=0}^{n_q-1}
	\gZZ_{pj;ql} \left( \nu''_{pq} \right)
\end{align}
where $\nu''_{pq}=2^{2n_q-4-p-q}\theta \Delta_\vphi^2$.
This Trotter steps consists of $n_q^2$ $\gZZ$ gates or $2 n_q^2$ CNOT gates. Since the interaction is not local, in this case the $\gZZ$ gates act on one qubit belonging to the qubit register allocated for the field at site $j$ and on one qubit belonging to the qubit register allocated for the field at site $l$.

The Trotter step corresponding to the $\phi^4$ interaction term is
\begin{align}
	\label{eq:trphi4}
	e^{-i\theta \tPhi_j^4}=&\lf[ \prod_{p=0}^{n_q-1}\prod_{q=0}^{p-1}\prod_{r=0}^{q-1}\prod_{s=0}^{r-1} \gZZZZ_{pj;qj;rj;sj}(\rho_{qprs})\rt] \nn\\
	& \times \lf[\prod_{p=0}^{n_q-1}\prod_{q=0}^{p-1} \gZZ_{pj;qj}(\eta_{pq})\rt]e^{i\xi}
\end{align}
where
\begin{align}
	\label{eq:z4gate}
	\gZZZZ_{pj;qj;rj;sj}(\rho)=e^{-i \rho \sigma^z_{pj} \sigma^z_{qj} \sigma^z_{rj} \sigma^z_{sj} }
\end{align}
and
\begin{align}
	\rho_{pqrs}&= \frac{3N_\vphi^4}{32} \frac{1}{2^{p+q+r+s}} \theta \Delta_{\vphi}^4\\ 
	\eta_{pq}&= \frac{N_\vphi^4}{16} \frac{1}{2^{p+q}} \left(1- \frac{1}{N_\vphi^2}-  \frac{1}{2^{2p+1}}- \frac{1}{2^{2q+1}} \right) \theta  \Delta_{\vphi}^4  \\
	\xi&= \left[\frac{(N_\vphi^2 - 1)^2}{48} -\frac{N_\vphi^4 - 1}{120}\right]\theta \Delta_{\vphi}^4
\end{align}
This step requires $n_q(n_q-1)(n_q-3)(n_q-3)/24$ four-qubit $\gZZZZ$ gates and $n_q(n_q-1)/2$ two-qubit $\gZZ$ gates.

\begin{table}
	\begin{tabular}{|c |c | c | c | c | c |}
		\hline \begin{tabular}{@{}c@{}}
			Operator
		\end{tabular}
		& $e^{-i \Phi \theta}$ & $e^{-i \Phi^2 \theta}$ & $e^{-i \Pi^2 \theta}$  & $e^{-i \Phi_j \Phi_k \theta}$ &  $e^{-i \Phi^4 \theta}$\\ [0.5ex]
		\hline \begin{tabular}{@{}c@{}}
			Number of \\ CNOTs
		\end{tabular}
		& 0 & $\qregsize^2 - \qregsize$ &
		$3\qregsize^2 - 3\qregsize$
		& $2 \qregsize^2$ & \begin{tabular}{@{}c@{}}
			$\frac{1}{4}\qregsize^4 - \frac{3}{2}\qregsize^3$ \\ $+\frac{15}{4} \qregsize^2 -\frac{5}{2} \qregsize $
		\end{tabular}  \\ [1ex]
		\hline	
	\end{tabular}
	\caption{CNOT gates count of the Trotter steps required for the implementation of the 
		$\phi^4$ evolution operator. All to all qubit connectivity is assumed.}
	\label{table:gate_counts}
\end{table}

Quantum simulation on near-term quantum devices is mainly limited by the two-qubit gate fidelities. The implementation of the Trotter step corresponding to the $\phi^4$ interaction term is computationally the most expensive one, since it requires $\O(n_q^4)$ of two-qubit gates. The number of CNOT gates for all Trotter steps relevant for the implementation of the $\phi^4$ evolution are summarized in \cref{table:gate_counts}. For comparison purpose, the $\gZZ$ and the $\gZZZZ$ gates are decomposed in two-qubit $CX$ (CNOT) gates and single-qubit $R^z$ rotations \cite{Welch2014},
\begin{align}
	\mathclap{Z}Z_{p;q}(\nu) =&CX_{p;q} R^z_q (\nu) CX_{p;q}, \\ 
	\gZZZZ_{p;q;r;s}(\rho)=&CX_{p;q}CX_{q;r}CX_{r;s}R^z_s (\rho) \nn\\
	& \times CX_{r;s}CX_{q;r}CX_{p;q}.
\end{align}
Since the number of Trotter steps is proportional to the lattice size, the computational cost of this algorithm scales linearly with $N$.

\section{State preparation}
\label{sec:state_preparation}

This section addresses the preparation of the ground state on qubits for both  normal and  broken-symmetry phases. Our method combines variational quantum circuits and adiabatic evolution, and it is flexible enough to allow tuning of different parameters to minimize circuit depth.  
To prepare broken-symmetry states, an interaction between the scalar field and an external field is introduced that explicitly breaks the $Z_2$ symmetry.
By properly choosing the strength of the external field as a function of time during the adiabatic process, the dual problems of degeneracy and broken-symmetry in the ground state are mitigated, as we discuss in \cref{ssec:adiabatic_prep_bs}.

The Hamiltonian $H$ employed for the quantum simulations of the $\phi^4$ model is given by \eqref{eq:Hdis1} with the lattice field operators replaced by the discrete field operators, as described in \cref{ssec:repr}. To prepare the ground state, we divide $H$ in two parts,
\begin{align}
	\label{Hphi4}
	H=H_{loc}+H_{c},
\end{align}
where
\begin{align}
	\label{eq:H0}
	H_{loc} &
	= \sum_{j=1}^N H_{loc,j} \\
	&= \sum_{j=1}^N \lf(
	\frac{1}{2} \Pi_{j}^2 + \frac{1}{2} \minit^2 \Phi_{j}^2
	+ \frac{\ginit}{4!} \Phi_{j}^4  + \finit \Phi_j \rt), \nn\\
	\label{eq:H1}
	H_{c} & =
	\sum_{j=1}^N \Bigg[
	\frac{1}{2}\sum_{e=0}^d \left(\Phi_{j+e } -\Phi_{j}\right)^2
	+ \frac{1}{2}\delta m^2\Phi_{j}^2
	\nn\\
	&  \quad\quad\quad \ + \frac{\delta \lambda}{4!} \Phi_{j}^4 + \delta f \Phi_j
	\Bigg],
\end{align}
where $\delta m^2=\latm_0^2-\minit^2$, $\delta \lambda=\latlam_0- \ginit$ and $\delta f=\latf_0-\finit$.
The Hamiltonian $H_{loc}$ is a sum of uncoupled local Hamiltonians $H_{loc,j}$ acting only at the lattice site $j$. The input parameters, $\minit^2$, $\ginit$ and $\finit$
should be chosen to ensure that the adiabatic evolution part of the state preparation is efficient, as we discuss in \cref{subsec:adiabatic}.
The first term in $H_c$  couples the fields at neighboring sites, while the last three terms in $H_c$ are local.

Our state preparation protocol consists of two parts.
\begin{enumerate}
	\item The ground state of $H_{loc}$ is prepared using variational circuits, as we describe in \cref{subsec:var_prep}.   It is a direct product of the ground state of $H_{loc,j}$ at each lattice site $j$,
	$\sket{\psi_\textrm{g}^{loc}}_j$:
	\begin{align}
		\label{eq:lattvarstate}
		\sket{\psi_{\textrm{g}}^{loc}}=\prod_{j=1}^N \otimes \sket{\psi_\textrm{g}^{loc}}_j.
	\end{align}
	\item The ground state of the full Hamiltonian is obtained by adiabatic evolution. The Hamiltonian $H_{c}$
	is turned on adiabatically. The system evolves under the time dependent Hamiltonian,
	\be
	\label{eq:time_dependent_H}
	H(s)=H_{loc} + \alpha(s) H_{c},
	\ee
	from $\sket{\psi_{\textrm{g}}^{loc}}$ to the ground state of the  Hamiltonian \cref{eq:Hdis1}.
	The time $t$ enters in \cref{eq:time_dependent_H} via the variable $s=t/T$, where $T$ is the total time of the adiabatic process and the function $\alpha(s)$ has boundary conditions $\alpha(0)=0$ and $\alpha(1)=1$ in the time interval $T$.
\end{enumerate}

\begin{figure*}[htb]
	\begin{center}
		\includegraphics*[width=5.5in]{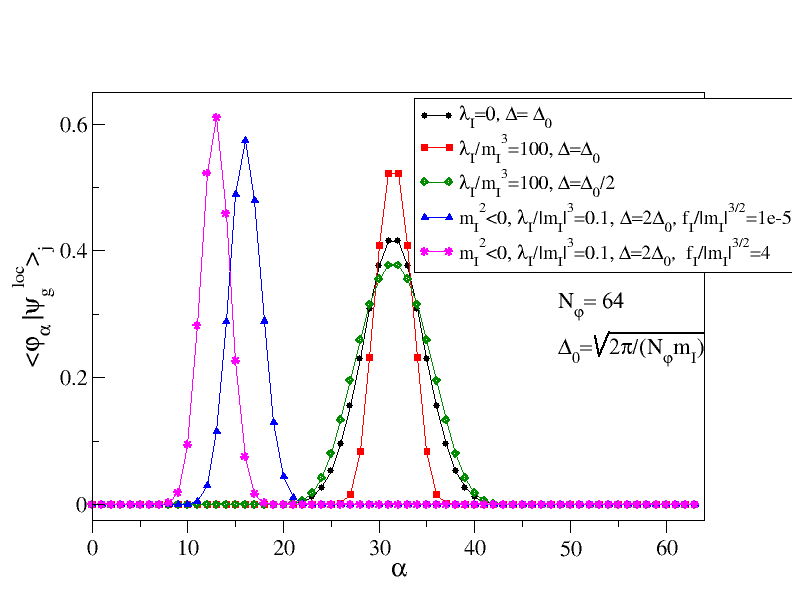}
		\caption{ Ground state wave functions, $\braket{\vphi_\alpha}{\psi^{loc}_{g}}_{j}$, of the local Hamiltonian $H_{loc,j}$ in \cref{eq:H0} represented on $n_q=6$ qubits ($N_\vphi=64$) vs the discretization index \edit{$\alpha \in \{0,1,\cdots,N_\vphi-1\}$} [see~\cref{eq:vphialpha}]. The black circles  
			illustrate the harmonic oscillator ground state ($\ginit=0$), which is a Gaussian. The red squares (green diamonds) illustrate the anharmonic oscillator ground state
			with strong interaction, $\ginit/\minit^3=100$, for a chosen discretization interval $\Delta=\Delta_0 \equiv \sqrt{2 \pi /(N_\vphi \minit)}$ ($\Delta=\Delta_0 /2$). The blue triangles illustrate the ground state for the Hamiltonian with negative mass-squqred and small external field. The magenta stars illustrate the ground state for the Hamiltonian with negative mass-squared and significant external field. These states are obtained by employing exact diagonalization.}
		\label{fig:localqstates}
	\end{center}
\end{figure*}

\begin{figure*}[htb]
	\begin{center}
		\includegraphics*[width=5.5in]{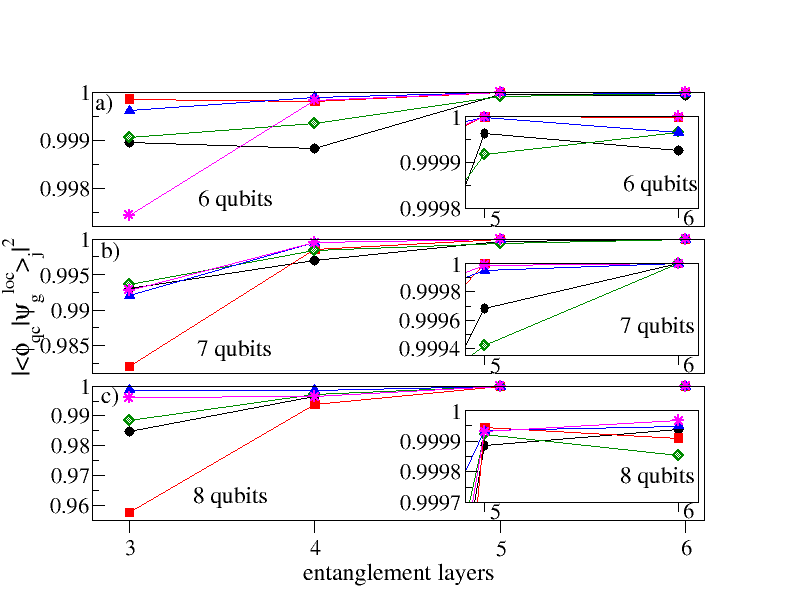}
		\caption{ (a) Fidelity of the quantum states illustrated in \cref{fig:localqstates} (same legend) prepared using an optimized quantum circuit vs the number of circuit's entanglement layers.
			(b) The same as in (a) when the quantum states
			are represented on $n_q=7$ qubits ($N_\vphi=128$).
			(c) The same as in (a) when the quantum states
			are represented on $n_q=8$ qubits ($N_\vphi=256$).
			The fidelity generally increases with the number of entanglement layers of the variational circuit. However the fidelity is not strictly monotonic since the circuit structure with an odd number of entanglement layer is different than that with an even number. For the cases with six, seven and eight qubits, the variational method reaches a fidelity $>0.9999$ with six entanglement layers.}
		\label{fig:variational_preparation}
	\end{center}
\end{figure*}

\subsection{Variational preparation of local states}
\label{subsec:var_prep}
The ground state $\sket{\psi_{\textrm{g}}^{loc}}_j$  of the local Hamiltonian $H_{loc,j}$  can be prepared accurately using short circuits on the $n_q$ qubits assigned to represent the field at the site $j$.  We propose a hardware-efficient circuit ansatz to prepare the local ground state using one- and two-qubit gates. The circuit parameters are determined using  optimization algorithms on classical computers, as discussed below.

First, the local wave function in the discrete field amplitude basis, $\braket{\vphi_\alpha}{\psi_{\textrm{g}}^{loc}}_j$ with \edit{ $\alpha \in \{0,1,\cdots,N_\vphi-1\}$}, is calculated on a classical computer. This  requires the diagonalization of a small size $N_\vphi \times N_\vphi$ matrix corresponding to the  Hamiltonian $H_{loc,j}$.

Second, parameterized quantum circuits are employed to produce 
$n_q$-qubit quantum states. 
We consider circuit ansatzes made by successive one-qubit and two-qubit layers. 
A one-qubit layer consists of one $R^y(\theta_i)$ rotation followed by one $R^z(\theta_j)$ rotation on every qubit. A two-qubit layer, which is responsible for introducing entanglement, consists of CZ gates acting on neighboring qubits. Qubit pairing in successive entanglement layers differs from each other and alternates. 
The quantum state $\ket{\phi_{qc}\left(\bt\right)}$ depends on $M$ rotation angles $\bt=\left(\theta_1,...,\theta_M\right)$ of the $R^y$ and $R^z$ single-qubit gates in the circuit.
Since, typically, $n_q$ is a small number ($6 \sim 8$), the state $\ket{\phi_{qc}\left(\bt\right)}$ can be computed on a classical computer without memory limitation problems, using packages such as \texttt{Cirq} \cite{cirq} or \texttt{Qiskit} \cite{qiskit}.

Third, the $M$ rotation angles, $\bt$, that parametrize the circuit are chosen so that the fidelity
\be
\label{eq:fid}
F(\bt)=\left|\braket{\phi_{qc}(\bt)}{\psi_{\textrm{g}}^{loc}}_j\right|^2
\ee
is as close to $1$ as possible. 
This can be accomplished, for example, by using the Covariance Matrix Adaptation Evaluation Strategy (CMA-ES) \cite{Hansen2001} for optimization on a classical computer.
CMA-ES is an iterative, genetic algorithm that generates a population of solutions at each iteration. The covariance matrix, calculated from a population subset with the largest values of $F(\bt)$, determines the population of solutions considered at the next iteration. The algorithm terminates when the best $F(\bt)$ of the population stops improving. The most difficult problem we encounter during the optimization of $F(\bt)$ is trapping at points of local maxima. We find that this problem can be avoided when CMA-ES runs with a large population of solutions. 

In \cref{fig:localqstates}, we show the ground states of the local Hamiltonian $H_{loc,j}$ represented on $n_q=6$ qubits for different Hamiltonian parameters calculated using exact diagonalization.
We are going to prepare these states by the parameterized circuits to demonstrate the efficiency of the variational preparation.
For illustration, we have chosen parameters representing different regimes, such as noninteracting, strong interacting with positive squared mass, negative squared mass with small external field strength and negative squared mass with a significant external field strength.
Since, as mentioned in \cref{ssec:repr} and discussed at large 
in~\cite{Macridin2021}, the discretization interval
$\Delta_\vphi$ in an interacting model can be tuned 
to optimize the performance of the algorithm, we present examples with
$\Delta_\vphi=\sqrt{2\pi/(N_\vphi\minit)},~1/2\sqrt{2\pi/(N_\vphi\minit)},~2\sqrt{2\pi/(N_\vphi\minit)}$. 
Note that the wave function representation on qubits 
depends significantly on   $\Delta_\vphi$. 
For example, the wave function plotted with red squares and the one plotted with green diamonds both correspond to the same Hamiltonian parameters $\ginit/\minit^3=100$ but the discretization interval of the latter is a factor of $2$ smaller.

The state preparation fidelity of our parametrized circuits is shown in \cref{fig:variational_preparation} as a function of entanglement layers, for $n_q=6$, $n_q=7$  and $n_q=8$ qubits. The target states are the one illustrated in \cref{fig:localqstates}.
For all  examples, the fidelity  of the local ground states is larger that $0.9999$ when at least six entanglement layers are used.  This is sufficient to accurately simulate a large lattice model.
For example, the fidelity to prepare the local ground state of a  lattice with $N=100$ sites is estimated to be $0.9999^{100} \approx 0.99$, which is comparable to the typical two-qubit gate fidelity ($\sim 0.995$) on NISQ devices \cite{Reagor2018,Arute2019,Jurcevic2021}. If a higher fidelity is needed,  more entanglement layers can be added to the variational circuit.

We do not encounter the difficulties seen in common variational quantum approaches such as the variational quantum eigensolver (VQE) \cite{O'Malley2016,Kandala2017} since the calculation of the quantum circuit required for state preparation is done on classical computers.
We do not require a global minimum and any solution with high enough fidelity ({\em i.e.} larger than the target accuracy) is acceptable.
The number of qubits $n_q$ needed to prepare the local wave function is not very large.
For example, $n_q=8$ is enough even for the strong interacting regime, since it can accommodate $\approx 200$ bosons per lattice site with great precision \cite{Macridin2021}.
Hence, the barren plateau issue ~\cite{McClean2018} is not a significant concern in our optimization. 
The optimization problem does not worsen when the system size is increased, since the optimized wave functions are local.
For the preparation of an $N$-site lattice wave function $\ket{\psi_\textrm{g}^{loc}}$  [\cref{eq:lattvarstate}], a quantum circuit running $N$ parallel $n_q$ quantum circuits  should be used. 

The variational circuit ansatz used here and 
constructed from $\mathrm{R_Y}$, $\mathrm{R_Z}$ and $\mathrm{CZ}$
gates is just a representative example.
Different hardware-efficient circuit ansantzes can be employed.
For example, the results of this section will be similar if one changes 
the $\mathrm{CZ}$ gate to the $\mathrm{CNOT}$ gate and the 
$\mathrm{R_Y}$ to the $\mathrm{R_X}$.

\subsection{Adiabatic evolution}
\label{subsec:adiabatic}

Our state preparation method relies on the adiabatic theorem~\cite{Born1928}, which relates the ground state of the interacting $\phi^4$ Hamiltonian $H$ [\cref{Hphi4}] to the ground state of the local Hamiltonian $H_{loc}$ [\cref{eq:H0}] under the action of the time dependent Hamiltonian $H(s=t/T)$ [\cref{eq:time_dependent_H}] for a sufficiently long time $T$.

There is a vast literature~\cite{Albash2018} addressing the necessary and sufficient conditions the adiabatic time  $T$ should fulfill.
A necessary condition is given by
\begin{align}
	\label{eq:adiabatic_condition} 
	&T \gg \frac {1}{\epsilon}
	\max_{s \in [0, 1]}
	\left|A_{m0}(s)\right|
	\text{\ \ for all $m \neq 0$, with} \\
	\label{eq:Adef}
	&A_{m0}(s)=\frac{ 
		\braket{ \evaE{m}{s}} { \dot{E}_0(s) }
	}
	{  \evaE{m}{s} - \evaE{0}{s}  },
\end{align}
where $\ket{\evaE{m}{s}}$ is the $m$-th instantaneous eigenstate of $H(s)$ satisfying $H(s)\sket{\evaE{m}{s}}=E_m(s)\sket{\evaE{m}{s}}$.
The system starts evolving from $\sket{\evaE{0}{0}} \equiv \sket{\psi_\textrm{g}^{loc}}$.
The dot denotes the derivative with respect to the $s$ variable, $\sket{\dot{E}_0(s)} \equiv \frac{d}{ds} \sket{\evaE{0}{s}}$, and $\epsilon=\norm{U(1)\ket{\psi_\textrm{g}^{loc}}-\ket{\psi_\textrm{g}^{trg}}}$ quantifies the difference between the state at the end of the  the adiabatic evolution $U(1)\ket{\psi_\textrm{g}^{loc}}$ and the target state $\ket{\psi_\textrm{g}^{trg}} \equiv \ket{\evaE{0}{1}}$. 
\Cref{eq:adiabatic_condition} is not a sufficient condition but provides a good estimate of $T$ for a large number of problems \cite{Amin2009,tong_prl_2007}.
The combination of this condition with the relation
\be
\label{eq:eedot}
\braket{ \evaE{m}{s}} { \dot{E}_0(s) }=\frac{\opmatrix{\evaE{m}{s}}{\frac{dH(s)}{ds}}{\evaE{0}{s}}}
{\evaE{m}{s} - \evaE{0}{s}}~~\text{for}~m \ne 0
\ee
implies that $T$ scales as the square of the minimum excitation gap.
When the excitation gap along the evolution path vanishes, as it does when the system passes through a critical region, the adiabatic process fails.

The condition \cref{eq:adiabatic_condition} is not always sufficient to ensure adiabatic evolution, typical examples where it fails being Hamiltonians with oscillatory terms.
A further, necessary condition for the validity of the adiabatic approximation \cite{tong_prl_2007} is given by
\begin{align}
	\label{eq:adiabatic_condition_b} 
	&T \gg \frac {1}{\epsilon}
	\max_{s \in [0, 1]}
	\left|\frac{d}{ds}A_{m0}(s)\right|
	\text{\ \ for all $m \neq 0$}.
\end{align}
In our case, this second adiabatic condition is relevant for the preparation of the broken-symmetry state, as we will discuss in \cref{ssec:adiabatic_prep_bs}.

The adiabatic time is shown to be proportional to the changing rate of the ground state wave function along the adiabatic path \cite{Boxio_pra_2010,HuaiXin_2013}, {\em i.e.} $T \propto \int_0^1 ds \norm{~\ket{\dot{E}_0(s)}~}$.
Intuitively we expect that,  the closer are the initial and the target wave functions, the smaller is the  overall changing rate of the ground state along the path, and implicitly the required adiabatic time. In fact, for both  normal and broken-symmetry phase preparations,  we observe a direct correlation between the adiabatic time  and the local overlap of the initial and the final wave functions. Namely, a larger overlap correlates with a shorter adiabatic time, as we discuss in \cref{sssec:ad_normalphase,ssec:adiabatic_prep_bs}.

The initial wave function is determined by the parameters $\minit$, $\ginit$ and $\finit$.
In the next sections we will explore the influence of these parameters on the adiabatic process. Since these parameters are adjustable in our algorithm, we will make recommendations for their choices.

\edit{The time dependence of the adiabatic process might significantly influence the adiabatic time. 
While for the preparation of normal phase states, we consider only adiabatic paths with linear time dependence, for the preparation of broken-symmetry states, we propose an adiabatic path with an exponential time dependence. This choice of the time dependence will mitigate the complications caused by the degeneracy of the ground state, as we discuss in~\cref{ssec:adiabatic_prep_bs}.}

\begin{figure*}[tb]
	\begin{center}
		\includegraphics*[width=\linewidth]{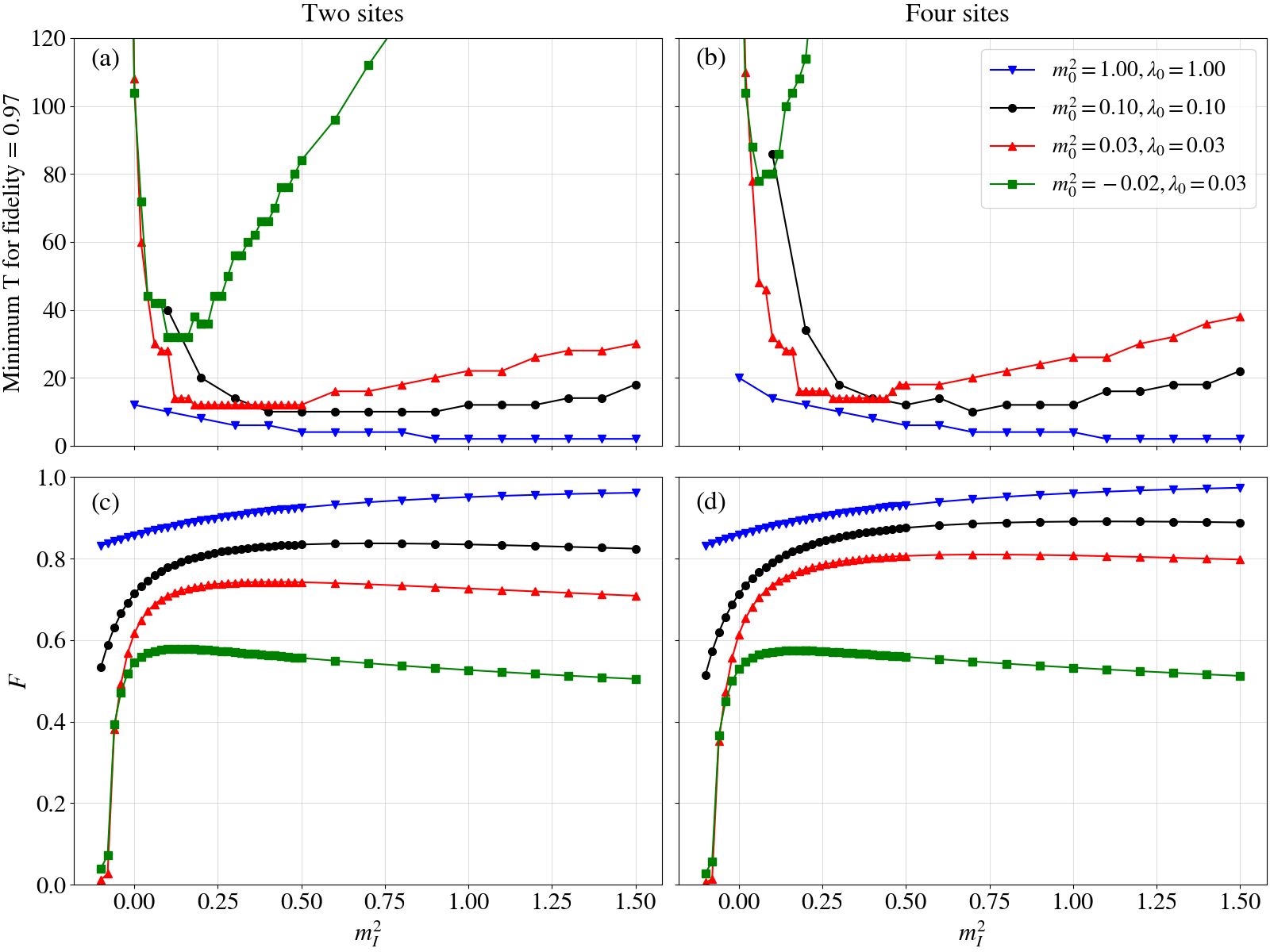}
		\caption{Adiabatic time $T$ required to prepare the $\phi^4$ normal state with $0.97$ fidelity vs $\minit^2$ for different values of the parameters $m_0^2$ and $\lambda_0$ for (a) two-site and (b) four-site lattices. $T$ sensitivity on $\minit^2$
			increases with decreasing $m_0^2$. (c) and (d) Local fidelity $F$, \cref{eq:localF}, measuring the overlap of the initial and the final local density
			matrices, vs $\minit^2$ for 
			two-site and, respectively, four-site lattices. The values of $\minit^2$ yielding the smallest $T$ yield 
			the largest $F_{loc}$.
			\edit{Note that the parameters and quantities displayed in this figure are dimensionless.}
		}
		\label{fig:normal_optimal_mu}
	\end{center}
\end{figure*}

\begin{figure*}[tb]
	\begin{center}
		\includegraphics*[width=\linewidth]{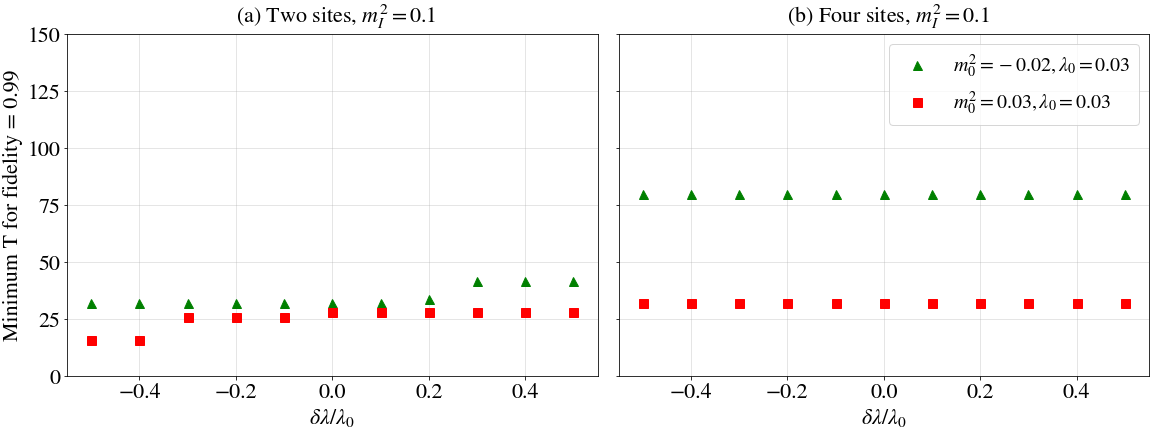}
		\caption{Adiabatic time $T$ as a function of $\delta \lambda$ for $\minit^2=0.1$,  different values  of the parameters $m_{0}^2$ and $\lambda_0$
			for (a) two-site and (b) four-site lattices. The adiabatic time $T$
			dependence on $\delta \lambda=\latlam_0-\ginit$ is weak.
			\edit{Note that the parameters and the adiabatic time $T$ displayed in this figure are dimensionless.}
		}
		\label{fig:normal_delta_ld0}
	\end{center}
\end{figure*}

\subsubsection{Preparation of normal phase states}
\label{sssec:ad_normalphase}

Finding the optimal adiabatic path for adiabatic evolution is difficult without the knowledge of the system's excitation spectrum. However, our goal in this section is less ambitious, and 
consists in investigating the effect of the  initial wave function $\ket{\psi_{\textrm{g}}^{loc}}$ (\cref{eq:lattvarstate})  on the adiabatic process.
For normal phase preparation  we  consider only adiabatic paths with linear time dependence, {\em{i.e.}} we take $\alpha(s)=s$ in \cref{eq:time_dependent_H}.

The normal phase of the $\phi^4$ model is characterized by a non-degenerate
ground state. The symmetry-breaking external field is unnecessary for ground state preparation in this case, and we set $\finit=0$ and $\delta f=0$ in \cref{eq:H0,eq:H1}. 
The dependence of  $\ket{\psi_{\textrm{g}}^{loc}}$ on $\minit$ and $\ginit$ can be understood by writing the local Hamiltonian (\cref{eq:H0}) as 
\begin{align}
	\label{eq:Hlocovm}
	\frac{H_{loc,j}}{\abs{\minit}}&=\frac{1}{2}\left(\frac{\Pi_j}{\sqrt{\abs{\minit}}}\right)^2+\text{sgn}(\minit^2)\frac{1}{2}\left(\sqrt{\abs{\minit}}{\Phi_j}\right)^2\\ \nonumber&
	+\frac{1}{4!}\frac{\ginit}{\abs{\minit}^3} \left(\sqrt{\abs{\minit}}\Phi_j\right)^4.
\end{align}
\noindent Up to a field amplitude scaling factor $\sqrt{\abs{\minit}}$ the eigenfunctions
of $H_{loc,j}$ are solely determined by the sign of $\minit^2$ and the ratio $\ginit/\abs{\minit}^3$, while $\abs{\minit}$ acts as a scaling factor for the energy.

The adiabatic evolution on small lattices is simulated on classical computers, using  the Trotterization method described in \cref{ssec:trotter}. We observe that $\minit$ has a strong influence on the adiabatic time and there is an optimal value that minimizes the adiabatic time. 
\Cref{fig:normal_optimal_mu} --(a) and (b) shows the adiabatic time $T$ needed to prepare
the ground state with $0.97$ fidelity as a function of $\minit^2$ for different values of the dimensionless bare parameters $m_0^2$ and $\lambda_0$ for  two- and four-site lattices.
The value of $T$ is especially sensitive to the mass parameter $\minit$  when  $m_0$ is small, a parameter regime relevant when taking the continuous limit $a \rightarrow 0$ (remember that $m_0=m_b a$).  We find that, in general, the optimal value of $\minit^2$ is larger than the bare mass $m_0^2$.

The dependence of the adiabatic time on $\minit$  can be understood by considering the strong influence of $\minit$ on the initial wave function, since $\sqrt{\abs{\minit}}$ acts as a scale factor for the field amplitude variable.
Two competing effects come into play to determine the optimal $\minit$. On one hand, the initial gap increases with increasing $\minit^2$, which favors the adiabatic process. On the other hand, increasing $\minit^2$ reduces the width of the initial wave function in the field amplitude basis. When the initial  wave function is too narrow, significant changes of the wave function are required during evolution, which increases the adiabatic time. 
In fact, for small lattices, we observe a direct correlation between the optimal adiabatic time and the local fidelity defined as~\cite{nielsen2002quantum}
\begin{align}
	\label{eq:localF}
	F_{loc}= {}_j\opmatrix{\psi_{\textrm{g}}^{loc}}{\rho_j}{\psi_{\textrm{g}}^{loc}}_j,
\end{align}
where $\rho_j = \prod_{k \neq j}\mathrm{Tr}_{k} \ketbra{\psi_\textrm{g}^{trg}}{\psi_\textrm{g}^{trg}}$ is the reduced density matrix at site $j$ of the system final ground state $\ket{\psi_\textrm{g}^{trg}} \equiv \sket{E_{0}(s=1)}$. In \cref{eq:localF}, $\mathrm{Tr}_{k}$ denotes the partial trace over a local basis at site $k$. The smallest $T$ is obtained for the values of $\minit^2$ which yield the largest values of $F_{loc}$, as can be seen by comparing the top panel and bottom panel of \cref{fig:normal_optimal_mu}.

Numerical simulations on small lattices reveal a weak dependence of the adiabatic time on the coupling strength $\ginit$ used to prepare the initial state.  \cref{fig:normal_delta_ld0} shows the time $T$ required to prepare the two- and four-site lattice ground state with fidelity $0.99$ as a function of  $\delta \lambda=\latlam_0- \ginit$. The parameters $m_0^2$ and $\lambda_0$ chosen in these  examples are small to accentuate the region where $T$ is sensitive to $\minit^2$. The chosen value of $\minit^2$ is close to the optimal one for $\delta \lambda=0$.  Particularly for the four-site lattice, the adiabatic time is nearly independent of $\ginit$.

\subsubsection{Preparation of broken-symmetry states}
\label{ssec:adiabatic_prep_bs}

Two issues related to the vanishing of the excitation gap must be addressed when preparing states in the broken-symmetry phase.
One is the potential crossing of the critical region during adiabatic evolution. The other is the degeneracy of the broken-symmetry ground state.
These problems can be overcome by introducing a time-dependent external field that couples to the scalar field during the adiabatic evolution.
The linear term in the Hamiltonian breaks the symmetry $\phi\to-\phi$ and, hence, the degeneracy of the ground state.
The field strength also controls the energy gap between the ground and first excited state.
By dividing the adiabatic evolution into two steps, we can focus on and discuss the gap and degeneracy problems independently.
The first adiabatic evolution starts from the ground state of the local Hamiltonian [\cref{eq:H0}]  and ends in the ground state of the full Hamiltonian with a finite external field.
The second adiabatic evolution starts from the terminus of the first one and ends when the external field is brought to near vanishing values (of the order of the desired error). 
Our notation is such that the external field changes from \fzero to \fone during the first stage of the evolution and from \fone to \ftwo in the second stage.

\paragraph{First adiabatic path. Avoiding the critical region. (\fzero $\longrightarrow$ \fone)}
\label{pg:efield}

\begin{figure*}[tb]
	\centering
	\includegraphics[width=\linewidth]{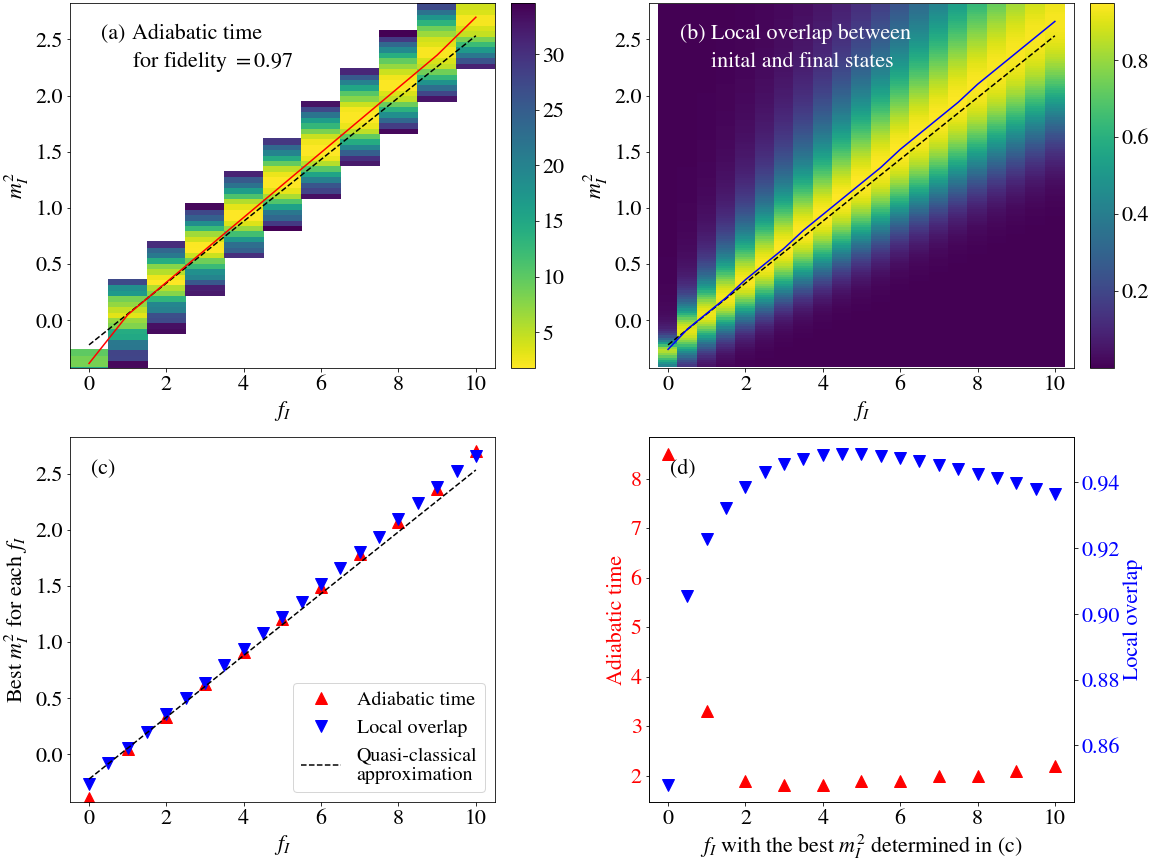}
	\caption{
		Diagnostics on the adiabatic evolution algorithm:
		(a) Time $T$ to reach a	fidelity of 0.97 as a function of initial $\minit^2$ and external field strength;
		(b) the local overlap in the same coordinate space;
		(c) $\minit^2$ which minimizes the adiabatic time and maximizes the local overlap for fixed $\fzero$ vs $\fzero$;
		(d) adiabatic time $T$ and the local overlap vs $\fzero$ for $\minit^2(\fzero)$ minimizing $T$ and extracted from (c).
		(\edit{Dimensionless parameters}: $m_{0}^2 = -0.22, \lambda_0= \ginit = 0.1, \fone=0.0011$)
	}
	\label{fig:broken_symmetry_comparison_1}
\end{figure*}

Previous studies of quantum algorithms for $\phi^4$ field theory \cite{Jordan2012} proposed state preparation via adiabatic evolution starting from the ground state of a noninteracting Hamiltonian.
Since the noninteracting ground state belongs to the normal phase region of the phase diagram, preparing broken-symmetry states in this way implies crossing the critical region characterized by a vanishing excitation gap. This is problematic, since the adiabatic process requires a finite gap.
Here we avoid crossing the critical region by starting the adiabatic evolution from a broken-symmetry state.

The first adiabatic path starts from a local state coupled to an external field $\fzero$ and ends in the ground state of the $\phi^4$ model coupled to the external field $\fone$.
The initial state is the ground state of ~\cref{eq:H0} and is prepared variationally as described in~\cref{subsec:var_prep}.
The first adiabatic process here is described by \cref{eq:time_dependent_H} with linear time dependence, $\alpha(s)=s$, and by ~\cref{eq:H1} with $\delta f=\fone-\finit$.
At the end of the first adiabatic path, the term containing the nonlocal coupling between sites is fully switched on.
Since the ground state of the broken-symmetry phase  of the $\phi^4$ model in zero external field is doubly degenerate (or nearly double degenerate for finite size lattices) and well separated from the rest of the spectrum (as numerical simulations presented in~\cref{fig:gaps14} shows),  \fone can be chosen small enough such that  the low-energy spectrum of the system at the end of the first adiabatic path can be approximated by a coupled two-level system  (\emph{i.e.}  $\fone \sabs{\sum_j \ssandwich{0}{1}{\Phi_j}} \ll E_2-E_1$,  where $E_1$ and $E_2$ are the energies of the first and
second excited states, respectively).
This choice of $\fone$, while providing a significant gap during the first adiabatic path,  will allow us to investigate analytically the second adiabatic process where 
the external field is taken to vanishing values.

\begin{figure*}[tb]
	\centering
	{\includegraphics[width=\linewidth]{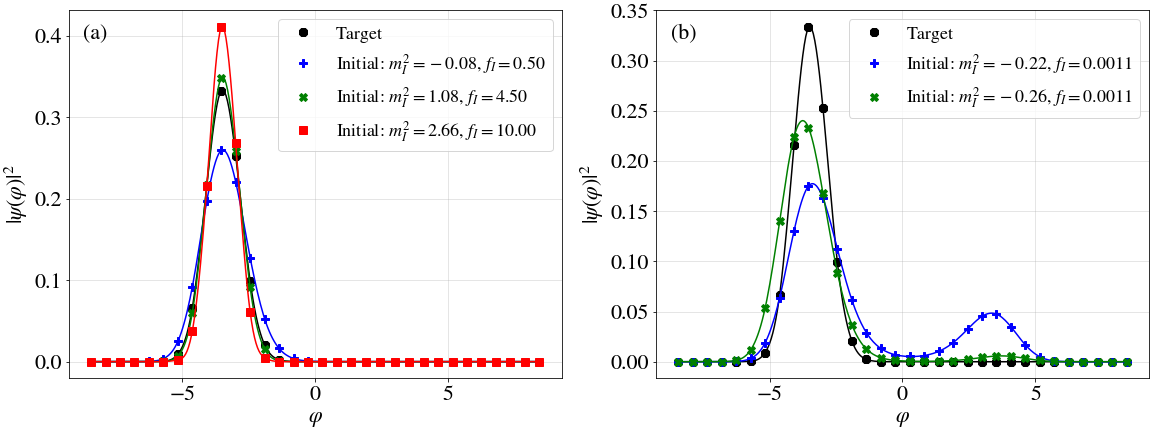} }
	\caption{
		Initial and target wave functions.
		The probability distribution along one of the field-amplitude coordinates (dimensionless) for different initial wave functions and the target wave function being the ground state of the full Hamiltonian.
		(a) $\minit^2$ dependence; (b) external field dependence.
		(\edit{Dimensionless parameters}: $m_{0}^2 = -0.22, \lambda_0= \ginit = 0.1, \fone=0.0011$)
	}
	\label{fig:broken_symmetry_wavefunction}
\end{figure*}

\begin{figure*}[tb]
	\centering
	{\includegraphics[width=\linewidth]{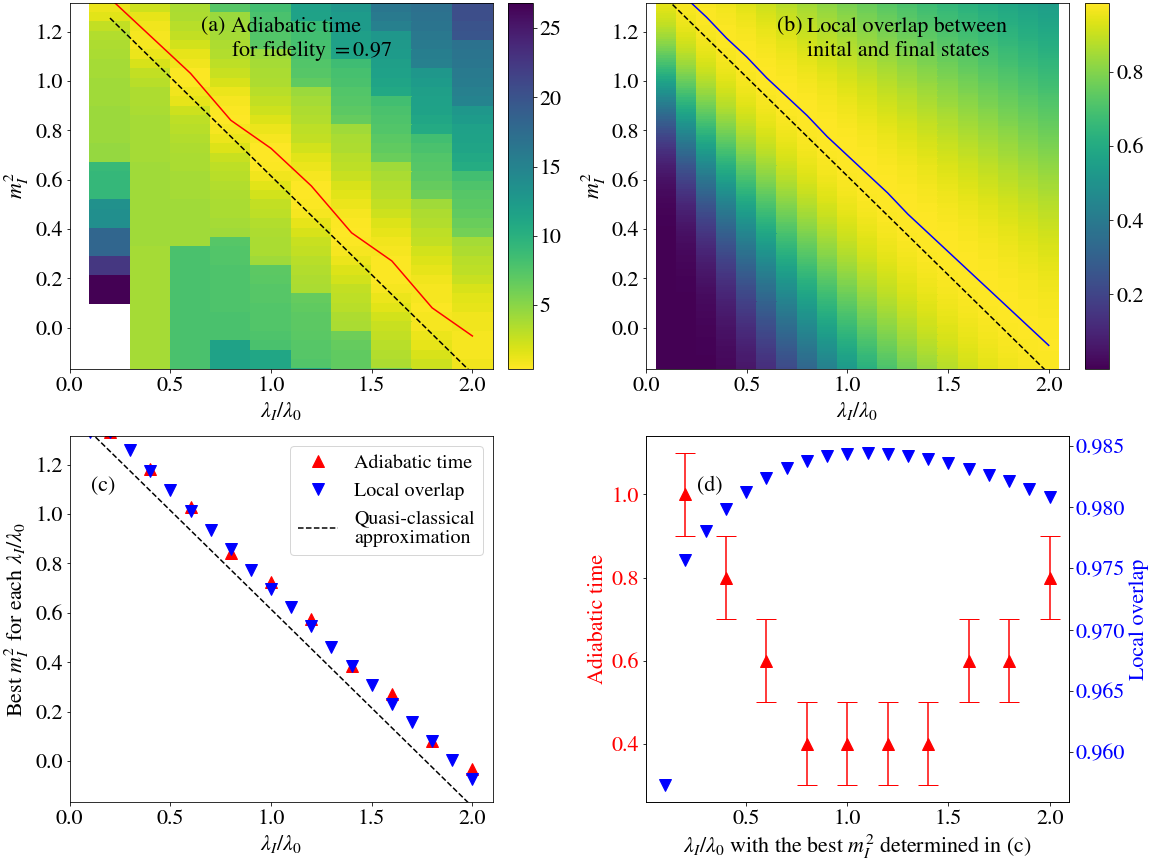} }
	\caption{
		Adiabatic evolution time and local overlap as a function of $\minit^2$ and $\ginit$ when
		$\finit$ is fixed to the value optimal for $\lambda_0= \ginit$. 
		(a) The time $T$ to reach a fidelity of 0.97 as a function of input parameters $\minit^2$ and initial $\ginit$; (b) The local overlap between the initial and the target wave function as a function of input parameters $\minit^2$ and initial $\ginit$; (c) $\minit^2$ yielding the shortest $T$ \edit{(red upper triangles)} and the largest local overlap \edit{(blue upside down triangles)} at fixed $\ginit$ vs $\ginit/\lambda_0$. (d) $T$ and local overlap vs $\ginit/\lambda_0$
		for $\minit=\minit(\ginit)$ yielding the shortest adiabatic time at fixed $\ginit$ extracted 
		from (c). The minimum adiabatic time and the maximum local overlap are found around $\ginit = \lambda_0$. (\edit{Dimensionless parameters}: $m_{0}^2 = -0.8, \lambda_0 = 0.6, \fone=0.0014$)
	}
	\label{fig:broken_symmetry_vary_g}
\end{figure*}

Our goal is to develop an algorithm that minimizes the adiabatic time by varying the algorithm input parameters, i.e., the initial magnitude field $\fzero$, the initial mass $\minit^2$ and the initial interaction strength $\ginit$.
Similar to the normal phase preparation, we find that the adiabatic time remains correlated with the local overlap between the initial state at $\fzero$ and the intermediate state at $\fone$. 
In the following, we investigate the adiabatic time dependence on input parameters through numerical simulations of two, three and four site lattices.

Simulations on small size lattices show that the shortest adiabatic times are correlated with the largest local overlap between the initial and final wave functions, similar
to the normal-phase case discussed in \cref{sssec:ad_normalphase}. An example  is shown in  \cref{fig:broken_symmetry_comparison_1} for the Hamiltonian parameters $m_{0}^2 = -0.22$, $\lambda_0= 0.1$, and $\fone=0.0011$. For this choice of the parameters, the coupling between sites (which is $1/2$ in the dimensionless units, see the first term in~\cref{eq:H1}) is the dominant term in the Hamiltonian.
The parameter $\ginit$ is fixed and equal to $\lambda_0$.
In \cref{fig:broken_symmetry_comparison_1} (a) we show the adiabatic time required to reach the fidelity $0.97$  while in \cref{fig:broken_symmetry_comparison_1} (b) we show the local overlap, both as a function of $\minit^2$ and $\fzero$.
The correlation can be clearly observed in 
\cref{fig:broken_symmetry_comparison_1}(c), where we plot the value of $\minit^2$ that yields the smallest $T$ (red upper triangles) and the largest local overlap (blue upside down triangles) for a fixed $\fzero$.
In \cref{fig:broken_symmetry_comparison_1}(d), we show the adiabatic time and local overlap
vs $\fzero$, with  $\minit^2=\minit^2(\fzero)$ extracted from \cref{fig:broken_symmetry_comparison_1}(c) to minimize $T$ at fixed $\fzero$. We find  that for the optimal adiabatic time, the external field $\fzero \gg \fone$ and the initial mass parameter $\minit^2>0 > \abs{m_0^2}$.
Similar investigations (with fixed $\ginit$) for other parameters of the Hamiltonian,  including smaller $m_{0}^2$ and $\lambda_0$ lead to similar conclusions (not shown).

The relation between the local overlap (and implicitly of the adiabatic time) and the parameters  $\minit^2$ and \fzero  can be understood by investigating the effect of these parameters on 
the initial wave function. The initial wave function's main peak  position and the peak's width  are dependent on both $\fzero$ and $\minit^2$. 
In \cref{fig:broken_symmetry_wavefunction}, we show the initial wave function distribution $\sabs{\braket{\vphi_\alpha}{\psi_g^{loc}}_j}^2$ for different input parameters $\minit^2$ and $\fzero$ together with  the local (at site $j$) probability distribution of the target wave function, defined as  $p(\vphi)_j=\opmatrix{\vphi_\alpha}{\rho_j}{\vphi_\alpha}$ [see \cref{eq:localF} for the definition  of $\rho_j$]. In the figure, the local field amplitude $\ket{\vphi_\alpha}$ is discretized to $N_\vphi=32$ points, corresponding to using five qubits for each site. 
\Cref{fig:broken_symmetry_wavefunction}(a) shows that the maximum overlap is obtained by choosing a value of $\fzero \gg \fone$, in agreement with the plot shown in \cref{fig:broken_symmetry_comparison_1}(d).
The initial wave function with the largest local overlap (green crosses) with the target wave function  is centered at the same location and has a similar width  as the target wave function (black dots).
By increasing (decreasing) $\fzero$ while keeping $\minit^2$ fixed, the wave function peak moves to 
the left (right). If the initial mass $\minit^2$ is increased (decreased) such  to keep the peak
aligned with the target wave function's peak, the wave function distribution becomes too 
narrow (wide) compared to the target one, as illustrated with red squares (blue pluses). Consequently the overlap decreases. The case where the external field $\fzero$ is chosen small, comparable to $\fone$, is shown in \cref{fig:broken_symmetry_wavefunction}(b). In this case, the local overlap is not optimal because the initial wave function exhibits a double-peak structure (green crosses and blue pluses).

An exhaustive, numerical search in a three dimensional space for the point $(\finit, \minit, \ginit)$ which minimizes
the adiabatic time is infeasible. Instead, we explore  the adiabatic time dependence  in the vicinity of the point $\left({\finit}_{0}, {\minit}_{0}, \lambda_0\right)$, where ${\finit}_{0}$ and ${\minit}_{0}$ are the initial external field and initial mass parameter which yield the shortest adiabatic time when $\ginit=\lambda_0$, as described in the example shown in \cref{fig:broken_symmetry_comparison_1}.  We find that modifying $\ginit$ in the vicinity this point does not reduce the adiabatic time.
For example, in \cref{fig:broken_symmetry_vary_g}, we show results for  Hamiltonian parameters $m_{0}^2 = -0.8$, $\lambda_0 = 0.6$, and $\fone=0.001$. 
We  keep $\finit={\finit}_{0}=4.0008$ fixed and vary $\ginit$ and $\minit^2$. The adiabatic time and the local overlap 
as a function of $\ginit/\lambda_0$ and $\minit^2$ are shown in \cref{fig:broken_symmetry_vary_g}(a) and (b), respectively. The minimum adiabatic time correlates with the maximum local overlap as in the previous example, as can be seen from \cref{fig:broken_symmetry_vary_g}(c) where the $\minit^2$ which yields the shortest adiabatic time and $\minit^2$ which yields the largest local overlap for fixed $\ginit$ are shown. While the adiabatic time dependence  on $\ginit$ is not negligible, 
we find that the shortest adiabatic time occurs when $\ginit \approx \lambda_0$, as can be seen in  \cref{fig:broken_symmetry_vary_g}(d). Simulations for other Hamiltonian parameters (not shown),
lead to the same conclusion: varying the input parameter $\ginit$ in the vicinity of $\lambda_0$ 
does not decrease the adiabatic time.

In \cref{fig:broken_symmetry_comparison_1,fig:broken_symmetry_vary_g}, it appears that there is a nearly linear relation between $\fzero$ and the corresponding best $\minit^2$ for both the adiabatic time and local overlap. This could be explained by the fact that, in this case, the local overlap is  maximized when the potential energy minimum is the same for the initial and final Hamiltonians. An approximation of that relation can be obtained using the quasi-classical argument given in \cref{app:quasiclassical} and is shown as black dashed lines.

\paragraph{Ground state degeneracy. (\fone $\longrightarrow$ \ftwo)}
\label{pg:gsdeg}

The broken-symmetry phase of the $\phi^4$ model  is characterized by a
twofold degenerate ground state in the thermodynamic limit and a
non-zero value of the order parameter. In numerical simulations, the broken-symmetry ground state
$\ket{\psi_{a0}}$ can be 
obtained from the ground state $\ket{\psi_{a}\left(L,f\right)}$ of
a system of finite size $L$ coupled to an external field $f$
\begin{align}
	\label{eq:Hf}
	H(f)&=H+f\sum_{j=1}^N \Phi_j
\end{align}
\noindent by taking the limits
\be
\label{eq:bslimit}
\ket{\psi_{a0}}=\lim_{f \to 0} \lim_{L \to \infty}\ket{\psi_a\left(L,f\right)}.
\ee
\noindent  \Cref{eq:Hf} is the same as \cref{eq:Hdis1} with $f\equiv f_0$ and with the
field operators replaced by the discrete field operators.
The limiting order in ~\cref{eq:bslimit} is important. When $L$ is finite, the system in the absence of the external field is not truly degenerate. 
An arbitrary small external field can drive the system to a broken
symmetry state only after the limit $L \rightarrow \infty$.
When estimating the limit~\cref{eq:bslimit} numerically, $f$ should be decreased 
and $L$ increased subject to the condition
that  $\sabs{f\opmatrix{0}{\sum_j \Phi_j}{1}} \gg \Delta_0$, where
the ground state of the system in zero external field is $\ket{0}$, the first excited state is $\ket{1}$,
and the energy gap between them is $\Delta_0=E_1-E_0$.

The degeneracy of the ground state in the broken-symmetry phase is
a problem for the adiabatic process, since the vanishing gap implies long adiabatic times. However, the quadratic adiabatic time scaling of the vanishing gap problem can be improved to linear by choosing an appropriate time dependence of the external field during the adiabatic process, as discussed in this section.
We focus here on the adiabatic evolution in the vicinity of the final state
characterized by vanishing external field. In this region, we assume that  the nearly double degenerate ground state is well separated from 
the rest of the spectrum. The external field during our adiabatic process
is always small such that the second term in~\cref{eq:Hf} can be considered a small perturbation.

We denote the two low-energy states in the presence of the external field $f$  by $\ket{\psi_a\left(f\right)}$ and $\ket{\psi_b\left(f\right)}$. As described in~\cref{app:PT},
a perturbative analysis reveals that the 
difference between the ground state of the system coupled to the external field $f$, $\ket{\psi_a(f)}$, and the broken-symmetry state, $\ket{\psi_{a0}}$, is 
\be
\label{eq:dist_bs}
\norm{\ket{\psi_a\left(f\right)}-\ket{\psi_{a0}}}=\frac{1}{2}fB +O(f^2), 
\ee
\noindent where $B$, explicitly derived in~\cref{eq:app_B}, is a quantity independent of $f$ and dependent on the $\sum_j \Phi_j$  matrix elements coupling the low-energy and the high-energy states.  The gap dependence on $f$ is given by
(see ~\cref{eq:app_gap_bs})
\be
\label{eq:gap_bs}
\Delta_{ba} \equiv E_b-E_a=2fv+O(f^2),
\ee
\noindent where
\be
\label{eq:vdef_order}
v=\abs{\opmatrix{\psi_{a0}}{\sum_j \Phi_j}{\psi_{a0}}}
\ee
\noindent  is equal to the order parameter in the broken-symmetry phase (see also~\cref{eq:vdef}).
The wave function dependence on $f$ yields
\begin{align}
	\label{eq:b_df_a}
	\opmatrix{\psi_b\left(f\right)}{\frac{d}{df}}{\psi_a\left(f\right)}= 
	\frac{B}{2} +O(f),
\end{align}
\noindent as derived in~\cref{eq:app_bdfa}.

The adiabatic process 
starts  from $\ket{\psi_{ini}}$,  the ground state of
Hamiltonian~\cref{eq:Hf} with the external field $f(s=0)\equiv
\fone$. At the end of the adiabatic evolution the external field is $f(s=1)\equiv \ftwo$.
The  error associated with the broken-symmetry state preparation is
\begin{align}
	\label{eq:err_def}
	\veps=\norm{U(1)\ket{\psi_{ini}}-\ket{\psi_{a0}}},
\end{align}
\noindent where $U(1)\ket{\psi_{ini}}$ is the adiabatically prepared state  and $\ket{\psi_{a0}}$ is the broken-symmetry state (target state).
There are two contribution to $\varepsilon$. The first one
is  caused by the finite value of the final external field,  and using \cref{eq:dist_bs}
is given by
\begin{align}
	\label{eq:epsilon_f}
	\varepsilon_f=\norm{\ket{\psi_a(\ftwo)}-\ket{\psi_{a0}}}=\frac{1}{2}\ftwo
	B,
\end{align}
\noindent where $\ket{\psi_a(\ftwo)}$ is the ground state of Hamiltonian~\cref{eq:Hf} with $f=\ftwo$. 
The second contribution is the error of the adiabatic process defined as the difference between the adiabatically prepared state and the ground state when the external field is $\ftwo$,
\begin{align}
	\label{eq:epsilon_a}
	\varepsilon_{ad}=\norm{U(1)\ket{\psi_{ini}}-\ket{\psi_{a}(\ftwo)}},
\end{align}
\noindent The triangle inequality implies that the total error (defined by \cref{eq:err_def}) is bounded by the sum
of these two contributions
\begin{align}
	\label{eq:epsilons}
	\varepsilon \le \varepsilon_f+\varepsilon_{ad}.
\end{align}

The challenge to preparing the broken-symmetry state through adiabatic evolution can be understood by inspecting the adiabatic condition 
~\cref{eq:adiabatic_condition}. 
Employing~\cref{eq:gap_bs,eq:b_df_a}  the term $A_{10}$  reads
\begin{align}
	\label{eq:adiab_A10}
	A_{10}(s)=\frac{B}{4v}  \frac{1}{f}\frac{df}{ds}.
\end{align}
\noindent  When $\frac{1}{f}\frac{df}{ds}$ is large, $\abs{A_{10}}$ and, implicitly, the adiabatic time become large.

For example, a large adiabatic time, with quadratic dependence on accuracy,  is required 
for an  external field with linear time dependence
\begin{align}
	\label{eq:linear_f}
	f(s)= \fone-\delta_f s,
\end{align}
where $\delta_f= \fone-\ftwo$.
The magnitude of $A_{10}(s)$ at the end of the adiabatic process  becomes
\begin{align}
	\label{eq:A0lin1}
	\abs{A_{10}(s=1)}=
	\frac{B^2 \delta_f}{8v}  \frac{1}{\veps_f}=
	\frac{\veps_{s=0} \left( \veps_{s=0}-\veps_f\right)}{\Delta_{s=0}} \frac{1}{\veps_f},
\end{align}
where $\Delta_{s=0} = 2 \fone v$ is the initial gap of
this adiabatic path and $\veps_{s=0}=B \fone/2$ is the difference
between the initial state and the target state.
If one naively employs ~\cref{eq:adiabatic_condition} to estimate
adiabatic time and assumes $\veps_{ad}\approx \veps_f \approx \veps/2$,  the adiabatic time reads $T \approx \frac{B^2 \delta_f}{8v}  \frac{1}{\veps_{ad}\veps_f} \approx \frac{B^2 \delta_f}{2v}\frac{1}{\veps^2}$. 
However,  the second adiabatic condition~\cref{eq:adiabatic_condition_b}, should also be considered for assessing the validity
of the adiabatic approximation. In this case 
it predicts an even longer adiabatic time,
$T \approx \frac{1}{\veps_{ad}}\abs{\frac{dA_{10}}{ds}(s=1)}=\frac{B^3\delta_f^2}{16v \veps_{ad}\veps_f^2} \approx \frac{B^3\delta_f^2}{2v \veps^3}$, thus an adiabatic time scaling as $\O(\veps^{-3})$. 
Fortunately, in the perturbative region where
the system can be reduced to a two level effective model, one can do better than employing the adiabatic conditions~\cref{eq:adiabatic_condition,eq:adiabatic_condition_b} for estimating the adiabatic time. Analytical and numerical calculations,  presented in~\cref{sssec:linear_path}, find that the 
best adiabatic time is obtained when $\ftwo$
is taken small  such that 
$\veps_f \ll \veps_{ad} \approx \veps$. In this case (see also~\cref{eq:Tlin1})
\begin{align}
	\label{eq:Tlin}
	T \approx \frac{\pi B^2 \delta_f}{16 v}  \frac{1}{\veps^2} \approx \frac{\pi}{2} \frac{\veps_{s=0}}{\Delta_{s=0}} \left(\veps_{s=0}-\veps_f\right)  \frac{1}{\veps^2}.
\end{align}
\noindent Thus, for an external field with
linear time dependence  the required adiabatic time  scales as
$\O(\veps^{-2})$. This is an
increase by a factor of $\veps^{-1}$ when compared to  systems with 
non-vanishing excitation gaps.

Since the excitation gap vanishes at the end of the adiabatic path, an external field time dependence which slows towards the end of the adiabatic evolution is expected to improve the adiabatic process.
As  ~\cref{eq:adiab_A10} predicts, the adiabatic
time does not blow up when $\frac{1}{f}\frac{df}{ds}$ is kept in bounds. For example, if the external field decreases exponentially, $T$ 
scales as $\O(\veps^{-1}\ln(\veps^{-1}))$, which
is better then $\O(\veps^{-2})$ for a linearly decreasing field. Indeed, by choosing
\begin{align}
	f(s) = \fone \exp( -\gamma s),~\text{with } \gamma=\ln(\fone/\ftwo)=\ln(\veps_{s=0}/\veps_f),
\end{align}
\noindent  one has
\begin{align}
	\label{eq:A10_exp}
	\abs{A_{10}}=\gamma \frac{B}{4v}=\gamma \frac{\veps_{s=0}}{\Delta_{s=0}},~~\text{ and  }   ~~\frac{d}{ds}A_{10}=0.  
\end{align}

\noindent  The adiabatic conditions~\cref{eq:adiabatic_condition,eq:adiabatic_condition_b} predict  an adiabatic time 
$T \approx \veps_{ad}^{-1}\ln( \veps_{f}^{-1})$. Taking  $\veps_f \approx \veps_{ad} \approx \veps/2$ this implies
$T \approx \veps^{-1}\ln( \veps^{-1})$.
In fact, explicit analytical and numerical calculations for a two-level systems, presented in~\cref{sssec:exp_path}, find that 
\begin{align}
	\label{eq:Texp}
	T \approx \frac{B}{2v \veps_{ad}} \ln\left( \frac{B \ftwo}{2\veps_f} \right)=2 \frac{\veps_{s=0}}{\Delta_{s=0} \veps_{ad}}\ln\left( \frac{\veps_{s=0}}{\veps_f} \right).
\end{align}
\noindent  For a desired accuracy $\veps$, one can show that (see~\cref{eq:Texp1,eq:Texp2})
\begin{align}
	\label{eq:Texp_bounds}
	\frac{B}{2v \veps} \ln\left( \frac{B \ftwo}{2\veps} \right) \le T \le \frac{B}{v \veps} \ln\left( \frac{B \ftwo}{\veps} \right).
\end{align}

Note that $T$ depends on the ratio $B/v= 4\veps_{s=0}/\Delta_{s=0}$.
A small ratio $B/v$ implies that the wave function's dependence on the
external field is much weaker than the
gap's dependence on the  external field. 
For the small size lattices explored here, we find  that $B/v$ 
is $\sim 10^{-2}$ (using the dimensionless units defining the
Hamiltonian \cref{eq:Hdis1}) close to the critical region
and is decreasing rapidly  when moving further away from the critical region.
We conclude that, for the adiabatic preparation of the broken-symmetry states,
the farther away from the critical region the states are the smaller adiabatic time is needed.

\subsubsection{Adiabatic evolution input parameters selection}
Finding the optimal input parameters $\finit$, $\minit^2$ and $\ginit$  for the preparation of large size lattice states is challenging. We  propose a strategy that avoids very long quantum circuits that are infeasible for limited-coherence near-term quantum hardware. First, determine the optimal input  parameters for a small lattice, as discussed here. Then, use those values as the starting point in a search for the optimal input parameters in increasingly larger systems, guided by local overlap measuring. The local overlap can be obtained by employing the SWAP test method~\cite{buhrman_prl_2001}.
This strategy implies running multiple evolution circuits for systems smaller than the target one. However, the evolution time for these runs is close to optimal, implying relatively short circuits.

\section{Conclusion}
\label{sec:conclusion}
In this paper, we present a circuit implementation of the evolution operator
of the  $\phi^4$ lattice  Hamiltonian on a qubit quantum computer and an algorithm to prepare states in both the normal and broken-symmetry phases.
The implementation is efficient in its use of resources and leverages the adjustable parameters of the problem to produce high fidelity states.
It is suitable for near-term quantum computers.

The evolution operator is implemented using the Trotterization method. The scalar 
field is encoded on the qubits using the discretized field amplitude representation \cite{Macridin2021} and requires a small number of qubits per site, $n_q \approx 6 \sim 8$. This number of qubits 
is adequate for exponential precision in even the strongly interacting regimes.
The required number of qubits and gates per Trotter step scale with 
the lattice size $N$. The most computationally expensive part of the evolution is the 
Trotter step associated with the $\phi^4$ interaction, which requires a number of two-qubit gates
proportional to $N n_q^4$.

Our state preparation 
combines a variational approach with
adiabatic evolution. The $\phi^4$ Hamiltonian is split into two parts: a local Hamiltonian consisting of a sum of local terms (\emph{i.e} uncoupled terms describing interaction at each site) and an inter-site coupling Hamiltonian that is switched on adiabatically to restore the full Hamiltonian.
The adiabatic process starts from the ground state of the local Hamiltonian, which is determined variationally.

The ground state of the local Hamiltonian is prepared with high fidelity using short quantum circuits.
These quantum circuits consists of a few ($\lessapprox 6$) two-qubit entangling layers ($CZ$ in our example) and parameterized single-qubit gates. The circuit parameters are calculated on a
classical computer by optimizing the overlap of the circuit final state and the local Hamiltonian ground state obtained from exact diagonalization. Since the local Hamiltonian is a sum of uncoupled terms at each site, the circuit optimization problem is reduced to the optimization of a circuit with a
small number of qubits
($n_q \approx 6 \sim 8$)  
and is independent of the lattice size.

The local Hamiltonian contains the $\phi^4$ interaction and, for the preparation of the broken-symmetry phase, a coupling of the scalar field to an external field. The parameters defining the local Hamiltonian,
the initial mass $\minit$, the initial interaction coupling $\ginit$ and the initial external field $\finit$,
constitute the input of our algorithm and can be adjusted. The system's Hamiltonian is restored by the adiabatic evolution. The initial parameters can be optimized to decrease the adiabatic time necessary 
to reach the full Hamiltonian ground state.

Our numerical investigation on small lattices finds a correlation between the adiabatic time and 
the local overlap of the final wave function and the initial wave function. To reduce the adiabatic time,
the input parameters $\minit$, $\ginit$ and $\finit$, should be chosen such that the final and the 
initial states have a maximum local overlap. For  state preparation in the normal phase, we find a strong dependence of
the adiabatic time on $\minit^2$ and a weak dependence on $\ginit$. In this case, the optimal $\minit^2$ is positive and larger than the $\phi^4$ Hamiltonian mass parameter $\abs{m_0}^2$.
For state preparation in the broken-symmetry phase, we find that the optimal adiabatic time is
achieved when the adiabatic process  starts from the ground state of a local Hamiltonian with significant external field $\finit$ and a positive input mass parameter $\minit^2$.

The correlation between the adiabatic time and the local overlap allows us to use the overlap as a tool to optimize the initial parameters. We propose an iterative strategy for finding the optimal input parameters $\minit$, $\ginit$ and $\finit$, starting with the optimal values for small lattices and adjusting them in increasingly larger systems by maximizing the local overlap.

There are two main challenges associated with the preparation of the broken-symmetry states that are addressed in this paper.
The first is when 
the adiabatic evolution crosses the critical phase transition region for
an initial state in the symmetric phase. The second is the vanishing gap of the double degenerate broken-symmetry phase.
We avoid these challenges by coupling the system to an external field during the adiabatic evolution.
We propose an adiabatic process consisting of two steps.
In the first step, the adiabatic evolution starts from a broken-symmetry state prepared variationally by coupling the local Hamiltonian to a finite external field. During this step, the coupling term is switched on and the external field is decreased.
During the second step, the external field is decreased to vanishing values.
The error in the adiabatic process can be kept under control by choosing a linear decrease of the external field in the first step and an exponential decrease in the second step.

\section{Acknowledgements}
A. M. is partially supported by the DOE/HEP QuantISED program grant of the theory  consortium "Intersections of QIS and Theoretical Particle Physics" at Fermilab."
A.C.Y.L.\ and S.M.\ are partially supported by the DOE/HEP QuantISED program grant "HEP Machine Learning and Optimization Go Quantum", identification number 0000240323. 
This manuscript has been authored by Fermi Research Alliance, LLC under Contract No. DE-AC02-07CH11359 with the U.S. Department of Energy, Office of Science, Office of High Energy Physics.

\appendix
\section{Fourier transform gate}
\label{app:ft_gate}
In this appendix, we prove $\cref{eq:fftqft}$,
which expresses the discrete Fourier transform gate $\F_j$ in terms of the QFT gate and single qubit rotations.
We begin from the definition of the discrete Fourier operator in \cref{eq:fft}:
\begin{align}
	\F_j =
	\frac{1}{\sqrt{N_\vphi}}
	\sum_{\alpha,\beta=0}^{N_{\vphi}-1} 
	e^{i \frac{2 \pi}{N_\vphi} \lf(\alpha-\frac{N_{\vphi}-1}{2}\rt) \lf(\beta-\frac{N_{\vphi}-1}{2}\rt)}
	\ket{\vphi_\alpha}_{j}\bra{\vphi_\beta}_j
	.
\end{align}
Expanding the phase factor and inserting the identity operator $\openone = \sum_{\alpha=0}^{N_{\vphi}-1} \ket{\vphi_\alpha}_{j}\bra{\vphi_\alpha}_j$ leads to
\begin{align}
	\label{eqapp:ft_gate}
	\F_j = &
	\, e^{i\frac{ N_\vphi \delta^2 }{2 \pi}}
	\nn\\
	&
	\lf( \sum_{\alpha=0}^{N_{\vphi}-1} e^{-i \delta \alpha} \ket{\vphi_\alpha}_{j}\bra{\vphi_\alpha}_j \rt)
	\nn\\
	&
	\lf(\frac{1}{\sqrt{N_\vphi}} \sum_{\mu, \nu=0}^{N_{\vphi}-1}  e^{i \frac{2 \pi}{N_\vphi} \mu \nu} \ket{\vphi_\mu}_{j}\bra{\vphi_\nu}_j \rt)
	\nn\\
	&
	\lf(\sum_{\beta=0}^{N_{\vphi}-1} e^{-i \delta \beta} \ket{\vphi_\beta}_{j}\bra{\vphi_\beta}_j \rt),
\end{align}
where $\delta = \tfrac{(N_{\vphi}-1) \pi }{N_\vphi}$.
The first line is a phase factor, and it is relevant if we would like to implement a control Fourier transform gate.
The third line is a standard QFT gate \cite{nielsen2002quantum}.
The second and the fourth lines can be implemented as single-qubit $z$ rotation gates similar to \cref{eq:trdispl} such that
\begin{align}
	& e^{-i \delta \alpha}  \ket{\vphi_\alpha}_{j}\bra{\vphi_\alpha}_j \nn\\
	= & e^{-i \delta \sum_{q=0}^{n_q-1} \alpha_{qj} 2^{n_q-1-q}}  \ket{\vphi_\alpha}_{j}\bra{\vphi_\alpha}_j  \nn\\
	= &  e^{-i \delta \sum_{q=0}^{n_q-1}   \frac{2\alpha_{qj} - 1}{2} 2^{n_q-1-q}} e^{-i \delta \frac{N_{\vphi}-1 }{2}}  \ket{\vphi_\alpha}_{j}\bra{\vphi_\alpha}_j \nn\\
	= &  e^{-i  \frac{N_{\vphi} \delta^2}{2\pi}}  \prod_{q=0}^{n_q-1} e^{-i \delta   \frac{\oszf{qj}}{2} 2^{n_q-1-q}}  \nn\\
	= &  e^{-i \delta^2 \frac{N_{\vphi} \delta^2}{2\pi}}  \prod_{q=0}^{n_q-1} R^z_{qj}\lf(2^{n_q-1-q} \delta\rt).
\end{align}
In the second line of the above equation, we used the binary representation $\alpha_j = \sum_{q=0}^{n_q-1} \alpha_{qj} 2^{n_q-1-q}$ [\cref{eq:binary}].
Rewriting \cref{eqapp:ft_gate} with QFT and $R^z$ gates, we arrive at \cref{eq:fftqft}
\begin{align}
	\F_j = & e^{-i\frac{ N_\vphi \delta^2 }{2 \pi}} \prod_{q=0}^{n_q-1}R^z_{qj}\left(2^{n_q-1-q} \delta\right) ~\text{QFT}_j
	\nn\\
	& \times \prod_{q=0}^{n_q-1} R^z_{qj}\left(2^{n_q-1-q}\delta \right).
\end{align}

\section{Semiclassical derivation of the relation between the optimal initial mass and the optimal initial external field}
\label{app:quasiclassical}
Here we present a semiclassical argument
to explain the nearly linear relation between the best $\minit^2$ and \fone shown in \cref{fig:broken_symmetry_comparison_1,fig:broken_symmetry_vary_g}.
The adiabatic preparation is optimized when the local overlap between the initial ground state of $H(s=0)$ and the final ground state of $H(s=1)$ is close to maximum.
While we do not have an analytic form of the ground-state wave function to determine the local overlap, we can determine the minimum of the classical potential energy, which in our case approximately predicts the center of the wave function.
Hence, we find that, when the minima of the initial potential energy and the final potential coincide, the adiabatic protocol is close to optimal.
The potential energy along the adiabatic path is given by
\begin{align}
	V(s)
	= \sum_{j=1}^N \bigg[
	&
	\frac{\minit^2 + s \, \delta m^2}{2} \Phi_{j}^2
	+ \frac{s}{2}\sum_{e=0}^d \left(\Phi_{j+e } -\Phi_{j}\right)^2
	\nn\\
	&
	+ \frac{\ginit + s \, \delta \lambda}{4!} \Phi_{j}^4  
	+ \lf(\finit+ s \, \delta f \rt) \Phi_j
	\bigg].
\end{align}
Given $m_{0}^2 < 0$ and assuming translational symmetry, the minimum of the potential energy is determined by
\be
\lf. \frac{\partial {V(s)} }{\partial \Phi_j} \rt|_{\Phi_j = \Phi_0(s)} = 0,
\ee
where $\Phi_0  =  \Phi_1 = \cdots = \Phi_{N=1} = \Phi_0(s) $ is the location of the minimum.
By requiring the minimum to be at the same position in the beginning (s=0) and at the end (s=1), i.e.\, $\Phi_0(0) = \Phi_0(1)$, we can find a relation between $\minit^2$ and $\finit$ as shown in \cref{fig:broken_symmetry_comparison_1,fig:broken_symmetry_vary_g} as black dashed lines.
In particular  the following linear relation,
\be
\label{eq:predicted_f_mu}
\finit = \mp \sqrt{\frac{3! \abs{m_0^2}}{\lambda_0}} \lf(
\minit^2 + \frac{\ginit}{\lambda_0} \abs{m_0^2}
\rt)
+ \mathcal{O}(\fone),
\ee
is a good approximation.
We emphasize that while this approximated relation aligns with the observations we made in the numerical simulation, this does not mean that the system properties in the low-energy subspace can be explained by a semiclassical argument. At best, the semiclassical argument gives us an insight about where the wave function is confined by the semiclassical potential in the field-amplitude basis. The distribution and most of the properties of the wave function remain strongly influenced by quantum effects and cannot be inferred from the semiclassical potential.

\section{Adiabatic preparation of a broken-symmetry state in an effective two-level system}
\label{app:PT}
\newcommand{\Ta}{T^{(a)}}
\newcommand{\Tb}{T^{(b)}}

\newcommand{\Tmin}{T_{m}}
\newcommand{\Tmina}{T^{(a)}_{m}}
\newcommand{\Tminb}{T^{(b)}_{m}}
\newcommand{\oH}{H}
\newcommand{\oHo}{H}
\newcommand{\oHp}{\Phi}
\newcommand{\oHeff}{\oH_{\mathrm{eff}}}
\newcommand{\oW}{W}

\newcommand{\oHa}{H^{(a)}}
\newcommand{\oHb}{H^{(b)}}

\newcommand{\Ediffa}[2]{\Delta^{(a)}_{#1, #2}}
\newcommand{\Ediffb}[2]{\Delta^{(b)}_{#1, #2}}

\newcommand{\evaEo}[1]{E^{(0)}_{#1}}
\newcommand{\evaEs}[1]{E_{#1}}
\newcommand{\evaEa}[2]{E^{(a)}_{#1}(#2)}

\newcommand{\evaEb}[2]{E^{(b)}_{#1}(#2)}
\newcommand{\eveEb}[2]{\ket{\evaEb{#1}{#2}}}
\newcommand{\seveEb}[2]{\sket{\evaEb{#1}{#2}}}

In this appendix, we investigate the adiabatic condition for the preparation of the broken-symmetry states near the end of the path in section \ref{pg:gsdeg}. Using an effective two-level system to approximate a nearly degenerate subspace, we will derive the adiabatic condition and propose more efficient adiabatic paths.

We consider a finite system of size $L$ in the parameter regime corresponding to the broken-symmetry phase. The  two lowest energy states  $\ket{0}$ and $\ket{1}$ are nearly degenerate and well separated from the rest of the spectrum,
\begin{align}
	\Delta_0 \equiv E_1-E_0 \ll E_2-E_1 \equiv \Delta_1.
\end{align}
In the thermodynamic limit, the states $\ket{0}$ and $\ket{1}$ are degenerate, {\em i.e.}
\begin{align}
	\Delta_0(L) \xrightarrow[L \rightarrow \infty]{} 0.
\end{align}
The broken-symmetry state can be reached by coupling the system to an external field
\begin{align}
	\label{appeq:Hf_sep}
	H_f&=H+f \Phi, ~~\text{~~where~~}\\
	\Phi&=\sum_{j=1}^N \Phi_j,
\end{align}
and then taking the limits $L\rightarrow \infty$ followed by $f \rightarrow 0$. Perturbation theory can be applied in this limit.
We then apply a Schrieffer-Wolff transformation \cite{Shavitt1980} 
to obtain an effective two-level Hamiltonian.

\subsection{Review of Schrieffer-Wolff transformation}
\label{app:effective_model}
In this subsection, we will provide a quick review of the Schrieffer-Wolff transformation.
In general, we can separate the full Hamiltonian into the unperturbed part $\oHo $ and the leading-order perturbation $\oHp$ such that
\be
\oH_{f} = \oHo + f \oHp.
\ee
The unperturbed part $\oHo$ is twofold degenerate due to a double-well potential.
We consider the nearly degenerate subspace to be spanned by the states $\sket{0}$ and $\sket{1}$, each localized at one of the potential wells with an energy $\evaE{0}{f=0}$.

We carry out the perturbative calculation in this subspace using a Schrieffer-Wolff transformation, which constructs an effective Hamiltonian $\oHeff = e^{-S} \oH_f e^{S}$ decoupling the twofold degenerate subspace from the higher-energy eigenstates up to an arbitrary order of $f$. The generator $S$ can be  constructed iteratively using a canonical van Vleck formalism \cite{Shavitt1980}.

To derive our result, we separate $\oHp$ into a block diagonal piece $V_D$ and a piece $V_X$ that couples the degenerate subspace and the higher-energy subspace.   The generator $S$ has a series expansion
\be
S = \sum_{j=1}^{\infty} f^j S^{(j)}.
\ee
Applying the Baker-Hausdorff lemma, we can expand the transformed Hamiltonian into
\begin{align}
	\oHeff = & \,
	e^{-S} \lf(\oHo + f V_D \rt) e^{S} + e^{-S} f V_X e^{S}
	\nn\\
	= & \,
	\oHo' + [\oHo + f V_D, S] + \frac{1}{2} [[\oHo, S], S]
	\nn\\
	& + f V_X  + f  [V_X, S]
	+ \mathcal{O}(f^3)
	\nn\\
	= & \,
	\oHo
	+ f \lf(V_D +  V_X  + [\oHo, S^{(1)}] \rt)
	\nn\\
	&
	+ f^2 \Big(
	\frac{1}{2} [[\oHo, S^{(1)}], S^{(1)}] 
	+ [V_X, S^{(1)}] + [\oHo, S^{(2)}]
	\nn\\
	& \ \ \ \ \ \ \ \  + [V_d, S^{(1)}] \Big)
	+ \mathcal{O}(f^3).
\end{align}
To decouple the two subspaces, we pick
\begin{align}
	[\oHo, S^{(1)}] & =  - V_X,
	\\
	[\oHo, S^{(2)}] & = - [V_d, S^{(1)}].
\end{align}
We can use the resolvent operator technique to determine the generator such that
\begin{align}
	S^{(1)} \ket{m} &=  - R_{m} [\oHo, S^{(1)}] \ket{m},
\end{align}
where the resolvent operator is given by
\begin{align}
	R_{m} & = \sum_{n  \neq m} \frac{\ketbra{n}{n}}{\evaEs{m} - \evaEs{n}}.
\end{align}
This gives
\begin{equation*}
	S^{(1)} =
	\sum_{k=0,1}  \sum_{\gamma  \ge 2}
	\lf(
	\frac{\sandwich{\gamma}{k}{V_X}}{\evaEs{k} - \evaEs{\gamma}} \ketbra{\gamma}{k}
	- \frac{\sandwich{k}{\gamma}{V_X}}{\evaEs{k} - \evaEs{\gamma}} \ketbra{k}{\gamma}
	\rt)
	.
\end{equation*}

The effective Hamiltonian up to the second order is given by
\be
\oHeff = \oHo + f V_D + f^2 \frac{1}{2} [V_X, S^{(1)}].
\ee
In the nearly degenerate subspace, $\oHeff$ can be written as
\begin{align}
	\label{eqapp:effective_hamiltonian_general}
	\oHeff = &
	\sum_{j=0, 1}\evaEs{j} \ketbra{j}{j}
	+f \sum_{j, k =0,1} V_{jk} \ketbra{j}{k}
	\nn\\
	&
	+f^2 \sum_{j, k =0,1} \oW_{jk} \ketbra{j}{k}
	+ \mathcal{O}(f^3),
\end{align}
where
$
\oW_{jk} =
\sum_{\gamma \ge 2} \frac{\ssandwich{j}{\gamma}{\oHp} \ssandwich{\gamma}{k}{\oHp}}{2}
\lf(
\frac{1}{ \evaEs{\gamma} - \evaEs{0} }
+ \frac{1}{ \evaEs{\gamma} - \evaEs{1} }
\rt)
$
and 
$V_{jk} = \ssandwich{j}{k}{\oHp}$.
The first two terms in the equation are simply the projection of $\oHo + f \oHp$ in the nearly degenerate subspace. The last term comes from the perturbative treatment and can be understood as the virtual interaction between the two nearly degenerate states through the higher-energy states.

\subsection{Effective two-level model}
\label{ssec:two_model}
Ignoring the higher-order terms in \cref{eqapp:effective_hamiltonian_general}, we get
\begin{align}
	\label{eq:Heff}
	H_{\mathrm{eff}}&=\sum_{j=0,1} E_j \ket{j}\bra{j}+ f\sum_{j,k=0,1}V_{jk}\ket{j}\bra{k} \nn\\
	&+f^2\sum_{j,k=0,1}W_{jk}\ket{j}\bra{k}.
\end{align}
Note that $V_{00}=0$, and  $V_{11}=0$ since $\ket{0}$ and $\ket{1}$ have the full symmetry of the Hamiltonian $H$, while $\Phi$ breaks the $Z_2$ symmetry.
For convenience we will denote 
\begin{align}
	\label{eq:vdef}
	v=\abs{V_{01}}=\abs{V_{10}}. 
\end{align}
The implicit assumption made when applying the perturbation theory is that $f$ is small such that $f v \ll \Delta_1$.  We also assume that the system is large  such that
$\Delta_0 \ll f v$. We ignore terms of $\O(\Delta_0/(fv))$ in the following.

The effective Hamiltonian \eqref{eq:Heff} acts on the  two-dimensional space spanned by the states
$\ket{0}$ and $\ket{1}$. The following notation is convenient:
\begin{align}
	\label{eq:w0w1}
	s&=\frac{1}{2}\left[E_0+E_1+f^2\left(W_{00}+W_{11}\right)\right],\\
	\delta&=\frac{1}{2}\left[\Delta_0+f^2 \left(W_{11}-W_{00}\right)\right],\\
	t&=fV_{01}+f^2W_{01},\\
	\label{eq:Ddef0}
	D&=\frac{\delta}{\sqrt{\delta^2+|t|^2}},\\
	t&=\abs{t}e^{-i 2 f},\\
	\label{eq:tanth}
	\tan 2 \theta &= -\frac{\abs{t}}{\delta},\\
	\label{eq:costh}
	\cos^2 \theta&=\frac{1}{2}\left(1+D\right),~~\cos \theta=\sqrt{\frac {1+D}{2}},\\
	\label{eq:sinth}
	\sin^2 \theta&=\frac{1}{2}\left(1-D \right),~~\sin \theta=-\sqrt{\frac {1-D}{2}}
	.
\end{align}
The eigenstates of the Hamiltonian~\eqref{eq:Heff} can be written as
\begin{align}
	\label{eq:eigens_eff}
	\ket{\psi_a\left(f\right)}&=e^{-if}\cos\theta\ket{0}+e^{if}\sin\theta \ket{1}\\
	\ket{\psi_b\left(f\right)}&=-e^{-if}\sin\theta\ket{0}+e^{if}\cos\theta \ket{1},
\end{align}
while the corresponding energies are
\begin{align}
	\label{eq:eigene_eff}
	{\cal{E}}_{a}&=s - \sqrt{\delta^2+|t|^2} \\
	{\cal{E}}_{b}&=s + \sqrt{\delta^2+|t|^2}.
\end{align}
Defining the broken-symmetry states as
\begin{align}
	\ket{\psi_{a0}}&=\lim_{f \rightarrow 0} \ket{\psi_a\left(f\right)}=\frac{1}{\sqrt{2}}\left(e^{-if}\ket{0}-e^{if}\ket{1}\right),\\
	\ket{\psi_{b0}}&=\lim_{f \rightarrow 0} \ket{\psi_b\left(f\right)}=\frac{1}{\sqrt{2}}\left(e^{-if}\ket{0}+e^{if}\ket{1}\right),
\end{align}
the eigenstates of the system can be written as
\begin{align}
	\label{eq:psiafvsbs}
	\ket{\psi_a(f)}&=\frac{\left(\cos \theta-\sin \theta\right)}{\sqrt{2}}\ket{\psi_{a0}}+\frac{\left(\cos \theta+\sin \theta\right)}{\sqrt{2}}\ket{\psi_{b0}} \\
	\label{eq:psibfvsbs}
	\ket{\psi_b(f)}&=-\frac{\left(\cos \theta+\sin \theta\right)}{\sqrt{2}}\ket{\psi_{a0}}+\frac{\left(\cos \theta-\sin \theta\right)}{\sqrt{2}}\ket{\psi_{b0}}.
\end{align}

The quantity $D$ in \cref{eq:Ddef0} can be written as
\begin{align}
	\label{eq:Ddef}
	D&=fB-f^2AB+\O(f^3)+\O(\frac{\Delta_0}{fv}),
\end{align}
where
\begin{align}
	\label{eq:app_A}
	A&=\frac{\left(V_{01}W_{10}+V_{10}W_{01}\right)}{2v^2},\\
	\label{eq:app_B}
	B&=\frac{\left(W_{11}-W_{00}\right)}{2v},
\end{align}
are independent on the external field $f$.
The following quantities, relevant for the investigation of the adiabatic process, can be  written up to $\O(f^3)$ and $\O(\Delta_0/(fv))$ as
\begin{align}
	\label{eq:app_gap_bs}
	\Delta_{ba}&={\cal{E}}_{b}-{\cal{E}}_{a}= 2fv\left(1+fA\right)\\
	\label{eq:app_bdfa}
	\opmatrix{\psi_b\left(f\right)}{\frac{d}{df}}{\psi_a\left(f\right)}&=-\sin \theta \frac{d}{df}\cos \theta+\cos \theta \frac{d}{df}\sin \theta\\ \nonumber
	&=\frac{1}{2}B-fAB+\frac{1}{4}f^2B^3+\O(f^3)\\
	\label{eq:app_adfb}
	\opmatrix{\psi_a\left(f\right)}{\frac{d}{df}}{\psi_b\left(f\right)}&=-\opmatrix{\psi_b\left(f\right)}{\frac{d}{df}}{\psi_a\left(f\right)}\\
	\label{eq:app_fdist}
	\norm{\ket{\psi_a\left(f\right)}-\ket{\psi_{a0}}}&=\frac{1}{2}D=\frac{1}{2}fB-\frac{1}{2}f^2AB.
\end{align}

\subsection{Adiabatic evolution in a two-level system}
\label{ssec:adiab_two_model}

Here, we investigate the two-level system's evolution when the external field $f(s)$, depending on the variable $s=t/T$, changes during a time interval $T$ from the initial value $f(s=0)=f_i$ to the final value $f(s=1)=f_f$.
Note that this evolution corresponds to the second stage of the adiabatic process to prepare the broken-symmetry states in \cref{ssec:adiabatic_prep_bs} such that $f_i = \fone$ and $f_f = \ftwo$.
The initial wave function is the ground state of the system when $f=f_i$, {\em i.e. } $\ket{\psi_{ini}} \equiv \ket{\psi(s=0)} =\ket{\psi_a(f_i)}$. During the adiabatic evolution, the wave function is
\begin{align}
	\label{eq:psis}
	\ket{\psi(s)}=c_a(s)\ket{\psi_a\left[f(s)\right]}+c_b(s)\ket{\psi_b\left[f(s)\right]}.
\end{align}
where $\ket{\psi_a(f)}$ and $\ket{\psi_b(f)}$ are the instantaneous eigenstates of the Hamiltonian~\eqref{eq:Heff}
with external field $f(s)$. 

The state at the end of the adiabatic evolution is 
\begin{align}
	\ket{\psi_f}\equiv \ket{\psi(s=1)}= c_{af}\ket{\psi_a(f_f)}+c_{bf}\ket{\psi_b(f_f)}
\end{align}
where we denote $c_{af} \equiv c_a(1)$ and $c_{bf} \equiv c_b(1)$.
The adiabatic error is given by the difference between the system's state at the end of the adiabatic evolution and the ground state when $f=f_f$,
\begin{align}
	\epsilon_{ad}&=\norm{\ket{\psi_f}-\ket{\psi_a(f_f)}}=\sqrt{2\left(1-\abs{c_{af}}\right)} \\ \nonumber
	&\approx \abs{c_{bf}} + \O(\abs{c_{bf}}^4)
\end{align}
The error caused by the finite final field, defined as the difference between the eigenstate when $f=f_f$ and the broken-symmetry state, is obtained by applying  \cref{eq:app_fdist}:\begin{align}
	\label{eq:epsf_def}
	\epsilon_f \equiv \norm{\ket{\psi_a(f_f)}-\ket{\psi_{a0}}}= \frac{1}{2}B f_f +\O(f^2).
\end{align}

Employing \cref{eq:psiafvsbs,eq:psibfvsbs,eq:psis}, the adiabatically prepared state can be written as
\begin{align}
	\ket{\psi_f}&=\frac{\cos \theta_f\left(c_{af}-c_{bf}\right)-\sin \theta_f\left(c_{af}+c_{bf}\right)}{\sqrt{2}}\ket{\psi_{a0}}\\ \nonumber
	&+\frac{\cos \theta_f\left(c_{af}+c_{bf}\right)+\sin \theta_f\left(c_{af}-c_{bf}\right)}{\sqrt{2}}\ket{\psi_{b0}},
\end{align}
where $\cos \theta_f$ and $\sin \theta_f$ are given by \cref{eq:costh,eq:sinth} when $f=f_f$.
The total error for preparing the broken-symmetry state is
\begin{align}
	\epsilon \equiv \norm{\ket{\psi_f-\ket{\psi_{a0}}}} \approx \abs{c_{af} \epsilon_f+c_{bf}}.
\end{align}
The coefficients $c_a(s)$ and $c_b(s)$, which describe the adiabatic evolution, obey the differential equations
\begin{align}
	\label{eq:diffeq1}
	\frac{dc_a}{ds}&=-c_b(s)\frac{df}{ds}\opmatrix{\psi_a\left(f\right)}{\frac{d}{df}}{\psi_b\left(f\right)}e^{-i T \int_0^s  \Delta_{ba}(u) du  }\\
	\label{eq:diffeq2}
	\frac{dc_b}{ds}&=-c_a(s)\frac{df}{ds} \opmatrix{\psi_b\left(f\right)}{\frac{d}{df}}{\psi_a\left(f\right)}e^{i T \int_0^s  \Delta_{ba}(u) du  },
\end{align}
with the initial conditions $c_a(0)=1$
and $c_b(0)=0$. These equations can be solved numerically. Next we will present results for
different choices of the time dependence of $f$: (1) linear and (2) exponential.

\subsubsection{Adiabatic evolution with linear dependence of the external field}
\label{sssec:linear_path}

\begin{figure*}[htb]
	\begin{center}
		\includegraphics*[width=7in]{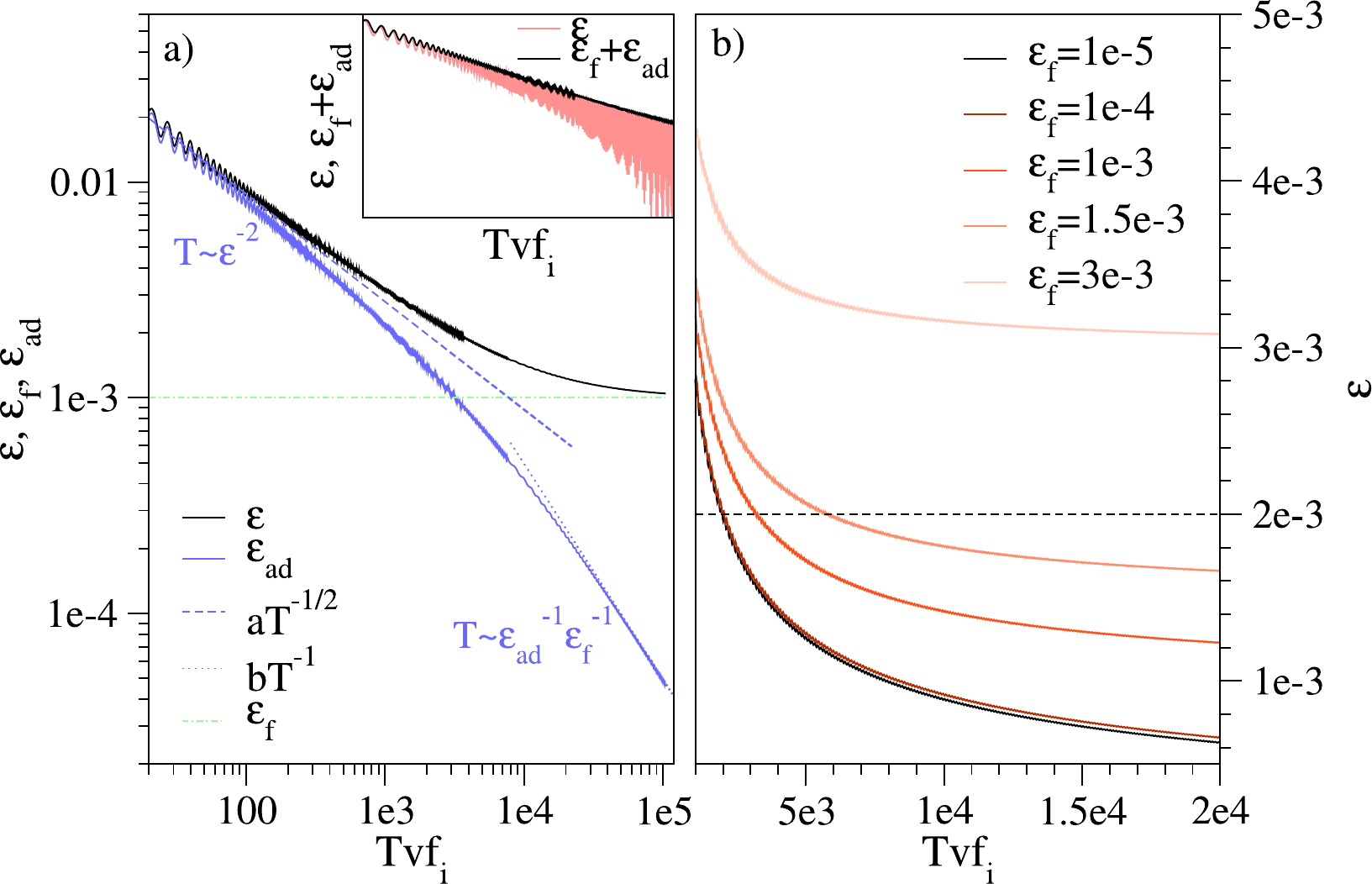}
		\caption{Errors for preparing the broken-symmetry state when the external field 
			has a linear time dependence in a two-level system with $B=0.2$, $v=10$ and $f_i=1$.
			a) Logarithmic scale. Errors $\epsilon$, $\epsilon_f$ and $\epsilon_{ad}$ vs the adiabatic time $T$ when $\epsilon_f=f_fB/2=10^{-3}$.
			The constants $a=B/4\sqrt{\pi\delta_f /v}$ and $b=\delta_f B^2/(8v\epsilon_f)$, see Eq.~\eqref{eq:Tlin1} and, respectively, Eq.~\eqref{eq:Tlin2}. Inset: The error $\epsilon$ (\edit{light} red line) has an oscillatory component as a function of $T$. The points of maximum are well approximated by $\epsilon_f+\epsilon_{ad}$ (black line).  b) The total error $\epsilon$ vs $T$ for different choices of the final external field $f_f=2\epsilon_f/B$. The adiabatic time for a fixed  $\epsilon$ decreases with decreasing $\epsilon_f$. The optimal adiabatic time is obtained in the parameter region where  $\epsilon_f \ll \epsilon \approx \epsilon_{ad}$.
			\edit{Note that the parameters and quantities displayed are dimensionless.}
			}
		\label{fig:linear_errors}
	\end{center}
\end{figure*}

In this scenario, the field's time dependence is
\begin{align}
	f(s)=f_{i}+(f_f-f_i)s=f_{i}-\delta_f s
\end{align}
where $\delta_f=f_i-f_f$.  \cref{eq:diffeq1,eq:diffeq2} reduce to
\begin{align}
	\label{eq:lin_ca}
	\frac{dc_a}{ds}&=-\frac{B}{2}\delta_fe^{-iTv\left(2 f_i s-\delta_f s^2\right)} c_b(s)\\
	\label{eq:lin_cb}
	\frac{dc_b}{ds}&=\frac{B}{2}\delta_fe^{iTv\left(2 f_i s-\delta_f s^2\right)}c_a(s).
\end{align}

At the end of this section we will present the  numerical solution to these equations. However,
in order to have an analytical estimate of the error dependence on the adiabatic time we proceed first by considering some simplifying approximations.  We will judge the accuracy of these approximations by comparing the approximate results with the exact numerical solution.

First, taking into account that $\abs{1-c_a(s)} \approx \epsilon_{ad}$,  \cref{eq:lin_cb}
can be written as
\begin{align}
	\label{eq:cb_approx}
	c_b(1) &= \frac{B}{2} \delta_f \int_0^1 ds   e^{i 2Tv \int_0^s  \left(f_i-\delta_f u\right) du  } + \O(\epsilon_{ad}^2) \\ \nonumber
	&=\frac{B}{2} \delta_f \int_0^1 ds e^{-i T v \delta_f \left( -2 \frac{f_i}{\delta_f} s+  s^2\right) } + \O(\epsilon_{ad}^2) 
	\\ \nonumber
	&=\frac{B}{2} \delta_f e^{i T v  \frac{f_i^2}{\delta_f}}\int_0^1 ds e^{-i T v \delta_f \left(s- \frac{f_i}{\delta_f}\right)^2 } + \O(\epsilon_{ad}^2)\\ \nonumber
	&=\frac{B}{2} \delta_f e^{i T v  \frac{f_i^2}{\delta_f}} \frac{1}{\sqrt{Tv\delta_f}}\int_{\sqrt{Tv\delta_f}
		\frac{f_f}{\delta_f}}^{\sqrt{Tv\delta_f}
		\frac{f_i}{\delta_f}} dy e^ {-i y^2}
	+ \O(\epsilon_{ad}^2)
\end{align}

The integral in Eq.~\eqref{eq:cb_approx}
can be expressed in terms of the error function  $\mathrm{Erf}(x)$, as
\begin{widetext}
	\begin{align}
		\label{eq:cberfs}
		c_b(1)=e^{i \frac{3}{4}
			\pi} \frac{B}{4}\sqrt{\frac{\pi \delta_f}{Tv}}e^{i T v  \frac{f_i^2}{\delta_f}}\left[\mathrm{Erf}\left(e^{i \frac{\pi}{4}}\sqrt{\frac{Tv}{\delta_f}} f_i\right)- \mathrm{Erf}\left(e^{i \frac{\pi}{4}}\sqrt{\frac{Tv}{\delta_f}} f_f\right)\right]+ \O(\epsilon^2).
	\end{align}
\end{widetext}
\noindent 
The error function expansion at small argument is
\begin{align}
	\label{eq:err_small}
	\mathrm{Erf}(e^{i\frac{\pi}{4}}x)=\sqrt{\frac{2}{\pi}} \left(1+i\right) x-\frac{2}{\sqrt{\pi}} e^{i\frac{3\pi}{4}}\frac{x^3}{3} +\O(x^5) 
\end{align}
\noindent while at large argument is
\begin{align}
	\label{eq:err_large}
	\mathrm{Erf}(e^{i\frac{\pi}{4}}x)=1-e^{-i x^2}\frac{1-i}{\sqrt{2 \pi}} \frac{1}{x}+\O(x^{-3}).
\end{align}
We distinguish two cases.

\paragraph*{Case $I$.} The adiabatic time satisfies
\be
\label{eq:Tcond1}
\frac{\delta_f}{v f_i^2} \ll T \ll \frac{\delta_f}{v f_f^2}.
\ee
In this case, the first error function in \cref{eq:cberfs} has a large argument while the second one has a small argument.
The adiabatic error is
\begin{align}
	\epsilon_{ad}&=\abs{c_b(1)}\\ \nonumber
	&\approx B\frac{\sqrt{\pi}}{4}\sqrt{\frac{\delta_f}{Tv}}\left(1-\sqrt{\frac{2}{\pi}} \sqrt{\frac{Tv}{\delta_f}}f_f\right)+\O(\frac{1}{T})\\ \nonumber
	& \approx B\frac{\sqrt{\pi}}{4}\sqrt{\frac{\delta_f}{Tv}}-B \frac{f_f}{ 2\sqrt{2}}.
\end{align}
This implies 
\begin{align}
	\left(\epsilon_{ad}+\frac{1}{\sqrt{2}} \epsilon_f \right)^2=\frac{\pi B^2 \delta_f}{16 T v}
\end{align}
and an adiabatic time scaling as
\begin{align}
	T \approx \frac{\pi B^2 \delta_f}{16 v} \frac{1}{\left(\epsilon_{ad}+\frac{1}{\sqrt{2}}\epsilon_f \right)^2}.
\end{align}
\Cref{eq:Tcond1} implies that this 
approximation is valid when
\begin{align}
	\frac{1}{4\left(\epsilon_{ad}+\frac{1}{\sqrt{2}}\epsilon_f \right)^2} \ll \frac{1 }{\pi \epsilon_f^2},
\end{align}
or equivalently when
\begin{align}
	\epsilon_f  \ll \frac{2}{\sqrt{\pi}-\sqrt{2} }\epsilon_{ad} \approx 5.58 \epsilon_{ad},
\end{align}
{\em i.e.} when the final external field is chosen small enough that  
\begin{align}
	\epsilon \approx \epsilon_{ad} \gg \epsilon_f.
\end{align}
In this case, the required adiabatic time scales
inversely proportionally with the squared accuracy, \begin{align}
	\label{eq:Tlin1}
	T \approx \frac{\pi B^2 \delta_f}{16 v} \frac{1}{\epsilon^2}.
\end{align}

\paragraph*{Case $II$.}
The adiabatic time satisfies
\be
\label{eq:Tcond2}
T \gg \frac{\delta_f}{v f_f^2}=\frac{\delta_fB^2}{4\epsilon_f^2}.
\ee
In this case, both error functions appearing in Eq.~\eqref{eq:cberfs} have a large argument. The adiabatic error can be approximated by
\begin{align}
	\epsilon_{ad}=\abs{c_b(1)} 
	\approx \frac{\delta_f B}{4 Tv f_f} =  \frac{\delta_f B^2}{8 Tv \epsilon_f},
\end{align}
which implies 
\begin{align}
	\label{eq:Tlin2}
	T \approx \frac{\delta_f B^2}{8 v \epsilon_f \epsilon_{ad}}.
\end{align}
\Cref{eq:Tcond2} implies that this 
approximation is valid when
\begin{align}
	\epsilon_f \gg  \epsilon_{ad}.
\end{align}
Making the assumption (numerically verified) that both $\epsilon_f$ and $\epsilon_{ad}$ are independent and the total error is the sum of  these two contributions, {\em i.e.} $\epsilon=\epsilon_f+\epsilon_{ad}$,  one gets
\begin{align}
	\label{eq:Tlin3}
	T > \frac{\delta_f B^2}{2v \epsilon^2 }.
\end{align}
By comparing the adiabatic time estimates for the two cases, \cref{eq:Tlin1,eq:Tlin2}, one concludes that, for a desired accuracy $\epsilon$, the choice of a large final external field (when $\epsilon \approx \epsilon_f \gg \epsilon_{ad}$) requires a larger $T$ than the choice of a small final external field (when $\epsilon \approx \epsilon_{ad} \gg \epsilon_f$).

\begin{figure*}[htb]
	\begin{center}
		\includegraphics*[width=7in]{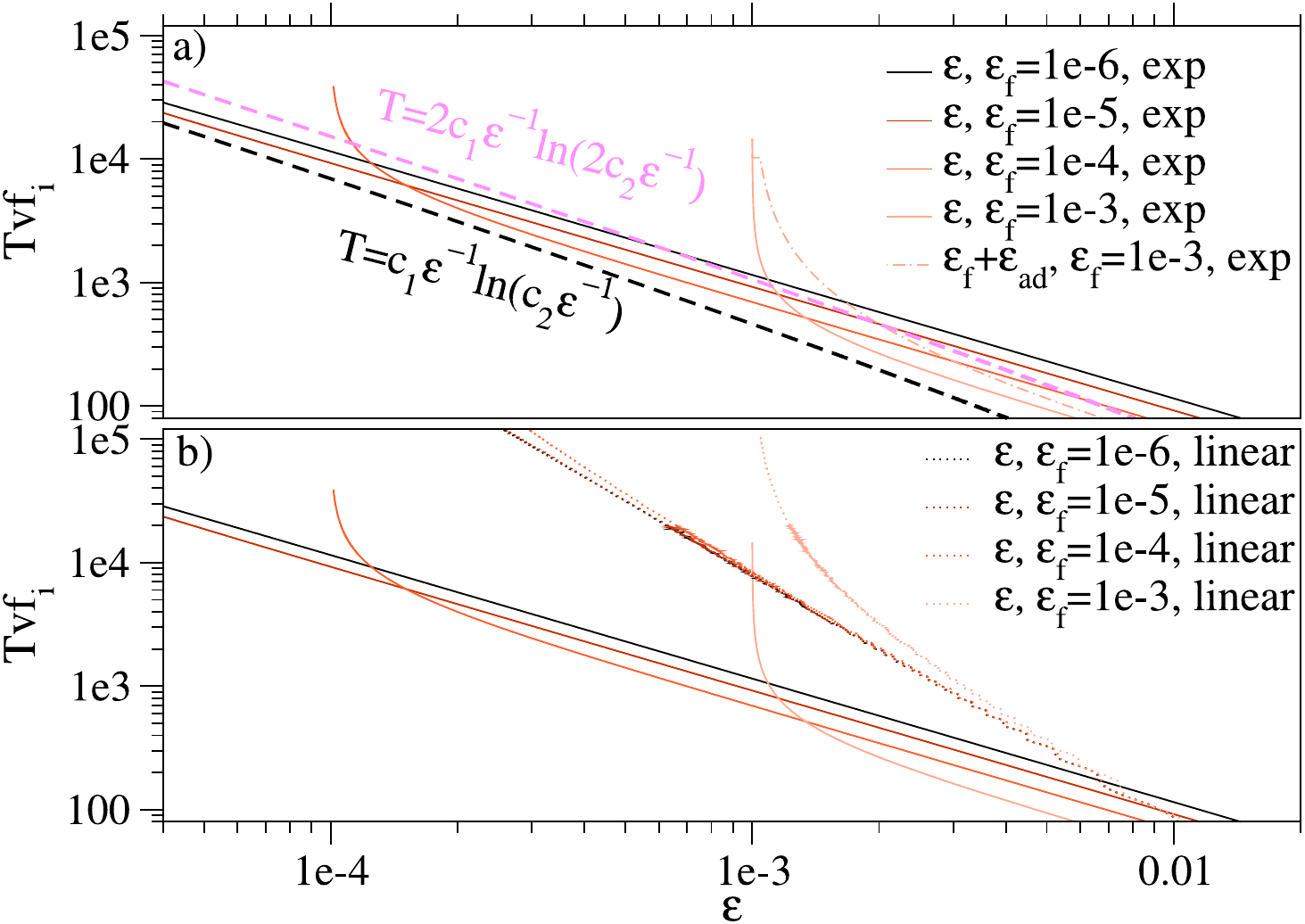}
		\caption{Adiabatic time to prepare the broken-symmetry state in a two-level system
			with $B=0.2$, $v=10$ and $f_i=1$. a) External field with exponential time dependence. $T$ vs $\epsilon$ for different values of the final external field $f=2 \epsilon_f/B$. When $\epsilon \gg \epsilon_f$, $T \approx \epsilon^{-1}\ln(\epsilon_f^{-1})$ in agreement with \cref{eq:Texp0}. The approximation $\epsilon \approx \epsilon_f +\epsilon_{ad}$ does not hold well when 
			$\epsilon$ approaches $\epsilon_f$, but the bounds for the adiabatic time given by \cref{eq:Texp1} and \cref{eq:Texp2} are valid (see dashed lines where $c_1=B/(2v)$ and $c_2 =B f_i/2$).
			b) Adiabatic time for the exponential path (\edit{solid} lines, same legend as in (a)) and for the linear path (dotted lines) vs accuracy. The adiabatic time for the exponential path scales as $\epsilon^{-1} \ln \epsilon^{-1}$ which is better than the $\epsilon^{-2}$ scaling for the linear path.
			\edit{Note that the parameters and quantities displayed in this figure are dimensionless.}
		}
		\label{fig:expandlin_errors}
	\end{center}
\end{figure*}

The errors  calculated by solving  \cref{eq:lin_ca,eq:lin_cb} numerically are shown in \cref{fig:linear_errors}. Since the right hand side of these differential equations
contains imaginary terms, the coefficients $c_{af}(T)$ and $c_{bf}(T)$ have an oscillatory component. As a consequence, $\epsilon$ and $\epsilon_{ad}$ display an oscillatory behavior, as can be seen from the inset. For our analysis of the numerical data, 
we consider the points where $\epsilon(T)$ reaches a local maxima.  As shown in the inset,  $\epsilon \approx \epsilon_f+\epsilon_{ad}$ at these points, {\em i.e.} the adiabatic error and the final field error can be considered as independent contributions to the total error. In \cref{fig:linear_errors} (a), we show the errors $\epsilon$, $\epsilon_f$ and $\epsilon_{ad}$ dependence on $T$ for a  case where the final external field yields $\epsilon_f=10^{-3}$. The numerical results are in agreement with the analytical analysis discussed earlier. At small $T$, where $\epsilon_{ad} \gg \epsilon_f$ the adiabatic time scales as $T \propto \epsilon^{-2}$, as predicted by \cref{eq:Tlin1}. At larger $T$ where $\epsilon_{ad} \ll \epsilon_f$ the adiabatic time scales as $T \propto \epsilon_f^{-1} \epsilon_{ad}^{-1}$, as predicted by \cref{eq:Tlin2}.

In ~\cref{fig:linear_errors} (b) we show
the total error $\epsilon$ vs $T$ for different values of the final external field. As our analytical analysis predicts, for a given $\epsilon$ (see for example the dashed \edit{black} line) the required adiabatic time decreases with decreasing  $\epsilon_f$. 

To conclude, we find that for an adiabatic process with linear time dependence of the external field, the adiabatic time scales as $\epsilon^{-2}$. 

\subsubsection{Adiabatic evolution with exponential dependence of the external field}
\label{sssec:exp_path}

In this scenario, the external field's time dependence is
\begin{align}
	f(s)=f_i e^{-\gamma s}~~ \text{ with}~~ \gamma=\ln\frac{f_i}{f_f}. 
\end{align}

\cref{eq:diffeq1,eq:diffeq2} reduce to
\begin{align}
	\label{eq:ca_deqexp}
	\frac{dc_a}{ds}&= -\frac{1}{2}\gamma f B e^{-i\frac{2Tv}{\gamma}\left(f_i-f\right)}c_b(s) \\
	\label{eq:cb_deqexp}
	\frac{dc_b}{ds}&=\frac{1}{2}\gamma f B e^{i\frac{2Tv}{\gamma}\left(f_i-f\right)}c_a(s)
\end{align}
Employing the approximation $\abs{1-c_a(s)} \approx \epsilon_{ad}$, one has
\begin{align}
	c_b(1) &\approx -\frac{1}{2}B \gamma \int_0^1 ds  f(s) e^{i 2Tv \int_0^s  f(u) du  } +\O(\epsilon_{ad}^2)\\ \nonumber
	&=\frac{iB\gamma}{4Tv} \left(e^{i 2Tv \int_0^1  f(s) ds} -1\right)+\O(\epsilon_{ad}^2)\\ \nonumber
	&=\frac{iB\gamma}{4Tv}\left(e^{i \frac{2Tv \delta_f}{\gamma} } -1\right)+\O(\epsilon_{ad}^2).
\end{align}
As for the linear adiabatic path case, $c_b(1)$ as a function of $T$ has an oscillatory behavior. For the $T$ values which yield local maxima of $\abs{c_b(1)}$ ({\em i.e.} $T=(2k+1)\pi\gamma/(2v\delta_f)$, with $k$ integer)
the adiabatic error is
\begin{align}
	\epsilon_{ad}=\abs{c_b(1)} \approx \frac{B\gamma}{2Tv} ,
\end{align}
implying
\begin{align}
	\label{eq:Texp0}
	T \approx \frac{B} {2v}\frac{1} {\epsilon_{ad}}\ln\frac{f_i}{f_f} =\frac{B} {2v}\frac{1} {\epsilon_{ad}}\ln\frac{Bf_i}{2\epsilon_f}.
\end{align}
Since $\epsilon_f, \epsilon_{ad} \le \epsilon$, a lower bound for $T$ is 
\begin{align}
	\label{eq:Texp1}
	T > \frac{B} {2v\epsilon}\ln\frac{Bf_i}{2\epsilon}.
\end{align}

For a desired accuracy $\epsilon$, it is possible to determine the optimal $\epsilon_f$ and $\epsilon_{ad}$ which minimize $T$ numerically.
To find an asymptotic estimate for $T$, we assume $\epsilon_f$ and $\epsilon_{ad}$ are independent contributions to the total error, {\em i.e.} $\epsilon \approx \epsilon_f+\epsilon_{ad}$. Then the optimal $T$ is smaller than or equal to
the one obtained for $\epsilon_f=\epsilon_{ad} \approx \epsilon/2$, {\em i.e.}
\begin{align}
	\label{eq:Texp2}
	T \le \frac{B} {v\epsilon}\ln\frac{Bf_i}{\epsilon}.
\end{align}
The bounds provided by \cref{eq:Texp1,eq:Texp2} show that  the time required for  adiabatic evolution scales as 
\begin{align}
	\label{eq:Texp3}
	T \propto \frac{1} {\epsilon}\ln\frac{1}{\epsilon},
\end{align}
which is an improvement compared to using  
an external field with linear time dependence, where $T \approx \epsilon^{-2}$ [see \cref{eq:Tlin1}].

The results obtained by solving  the
differential equations \eqref{eq:ca_deqexp} and \eqref{eq:cb_deqexp} numerically are shown in \cref{fig:expandlin_errors} (a). The adiabatic time satisfies \cref{eq:Texp0}. The approximation $\epsilon \approx \epsilon_f+\epsilon_{ad}$ does not hold well in the region where $\epsilon$ approaches $\epsilon_f$ (see the \edit{solid and dash-dotted lines with the lightest color corresponding to $\epsilon_f=10^{-3}$}). However, the inequalities \eqref{eq:Texp1} and \eqref{eq:Texp2} are true.
In ~\cref{fig:expandlin_errors} (b) we compare the adiabatic time for the linear adiabatic path (dotted lines) with the one corresponding to the exponential adiabatic path (\edit{solid} lines).
As \cref{eq:Tlin1,eq:Texp3} predict, the adiabatic time for the exponential case is much smaller than that for the linear case.

\bibliographystyle{apsrev4-2}
\bibliography{bib}

\end{document}